\documentclass[twocolumn]{aastex631}

\usepackage{graphicx}
\usepackage[caption=false]{subfig}	
\usepackage{longtable}
\usepackage{latexsym}
\usepackage{amsmath}
\usepackage{txfonts}
\usepackage{ulem}
\submitjournal{ApJ}

\shorttitle{Dark Galaxy Chain}
\shortauthors{J\'ozsa et al.}

\newcommand{\HIe}{H\,{\textsc i}}
\newcommand{\HI}{\textrm{H}\,{\textsc i}} 
\newcommand{\IDAVIE}{\textit{iDaVIE}}

\begin{document} 

\title{The detection of a massive chain of dark \HI\ clouds in the GAMA G23 Field}

\author{Gyula I. G. J\'ozsa}
\affiliation{Max-Planck-Institut f\"ur Radioastronomie, Radioobservatorium Effelsberg,
Max-Planck-Stra{\ss}e 28, 53902 Bad M\"unstereifel, Germany}
\affiliation{Department of Physics and Electronics, Rhodes University, PO Box 94, Makhanda, 6140, South Africa}
\affiliation{South African Radio Astronomy Observatory, 2 Fir Street, Black River Park, Observatory, Cape Town, 7925, South Africa}
\thanks{E-mail:gjozsa@mpifr-bonn.mpg.de} 

\author{T.H. Jarrett}
\affiliation{Dept.\ of Astronomy, Univ.\ of Cape Town, Private Bag X3, Rondebosch 7701, South Africa} 
\affiliation{Western Sydney University, Locked Bag 1797, Penrith South DC, NSW 1797, Australia}
\thanks{E-mail:jarrett@ast.uct.ac.za} 

\author{Michelle Cluver}
\affiliation{Centre for Astrophysics and Supercomputing, Swinburne University of Technology, John Street, Hawthorn, 3122, Australia  } 
\affiliation{Department of Physics and Astronomy, University of the Western Cape, Robert Sobukwe Road, Bellville, 7535, Republic of South Africa}
\thanks{E-mail:mcluver@swin.edu.au}

\author{O. Ivy Wong}
\affiliation{CSIRO Space \& Astronomy, PO Box 1130, Bentley, WA 6102, Australia}
\affiliation{International Centre for Radio Astronomy Research (ICRAR), University of Western Australia, Crawley, WA 6009, Australia}
\affiliation{ARC Centre of Excellence for Astrophysics in Three Dimensions (ASTRO 3D), Australia}

\author{Okkert Havenga}
\affiliation{Dept.\ of Astronomy, Univ.\ of Cape Town, Private Bag X3, Rondebosch 7701, South Africa} 

\author{H. F. M. Yao}
\affiliation{Department of Physics and Astronomy, University of the Western Cape, Robert Sobukwe Road, Bellville, 7535, Republic of South Africa}

\author { L. Marchetti}
\affiliation{Dept.\ of Astronomy, Univ.\ of Cape Town, Private Bag X3, Rondebosch 7701, South Africa}

\author{E.N. Taylor}
\affiliation{Centre for Astrophysics and Supercomputing, Swinburne University of Technology, John Street, Hawthorn, 3122, Australia  } 

\author{Peter Kamphuis}\affiliation{Ruhr University Bochum, Faculty of Physics and Astronomy, Astronomical Institute, 44780 Bochum, Germany }  

\author{Filippo M. Maccagni}\affiliation{INAF - Osservatorio Astronomico di Cagliari, Via della Scienza 5, I-09047 Selargius (CA), Italy }  

\author{Athanaseus J. T. Ramaila}\affiliation{South African Radio Astronomy Observatory, 2 Fir Street, Black River Park, Observatory, Cape Town, 7925, South Africa}  

\author{Paolo Serra}\affiliation{INAF - Osservatorio Astronomico di Cagliari, Via della Scienza 5, I-09047 Selargius (CA), Italy }  

\author{Oleg M. Smirnov}\affiliation{Department of Physics and Electronics, Rhodes University, PO Box 94, Makhanda, 6140, South Africa}
\affiliation{South African Radio Astronomy Observatory, 2 Fir Street, Black River Park, Observatory, Cape Town, 7925, South Africa}

\author{Sarah V. White}\affiliation{Department of Physics and Electronics, Rhodes University, PO Box 94, Makhanda, 6140, South Africa}

\author{Virginia Kilborn}
\affiliation{Centre for Astrophysics and Supercomputing, Swinburne University of Technology, John Street, Hawthorn, 3122, Australia  } 

\author{B.W. Holwerda}
\affiliation{Department of Physics and
Astronomy, University of Louisville, 102 Natural Science Building, Louisville, KY 40292, USA}

\author{A.M. Hopkins}
\affiliation{Australian Astronomical Optics, Macquarie University, 105 Delhi Rd, North Ryde, NSW 2113, Australia}

\author{S. Brough}
\affiliation{School of Physics, University of New South Wales, NSW 2052, Australia}

\author{K.A. Pimbblet}
\affiliation{E.A. Milne Centre for Astrophysics, University of Hull, Cottingham Road, Kingston-upon-Hull, Hull HU6 7RX, UK}

\author{Simon P. Driver}
\affiliation{International Centre for Radio Astronomy Research (ICRAR), University of Western Australia, Crawley, WA 6009, Australia}
\affiliation{School of Physics \& Astronomy, University of St Andrews, North Haugh, St Andrews, KY16 9SS, UK}

\author{K. Kuijken}
\affiliation{Leiden Observatory, Leiden University, PO Box 9513, 2300RA Leiden, the Netherlands}

 \begin{abstract}
We report on the detection of a large, extended \HI\ cloud complex in the GAMA G23 field, located at a redshift of $z\,\sim\,0.03$, observed as part of the MeerHOGS campaign (a pilot survey to explore the mosaicing capabilities of MeerKAT). The cloud complex, with a total mass of $10^{10.0}\,M_\odot$,
lies in proximity to a large galaxy group with $M_\mathrm{dyn}\sim10^{13.5}\,M_\odot$.
We identify seven \HI\ peak concentrations, interconnected as a tenuous `chain' structure, extending $\sim$400\,kpc from east-to-west, with the largest (central) concentration containing $10{^{9.7}}$\,$M_\odot$ in \HI\ gas distributed across 50\,kpc. The 
main source is not detected in ultra-violet, optical or infrared imaging.  The implied gas mass-to-light ($M_{\textrm{H}\textsc{I}}$/$L_\mathrm{r}$) is extreme ($>$1000) even in comparison to other `dark clouds'.
The complex has very little kinematic structure (110\,km\,s$^{-1}$), making it difficult to identify cloud rotation.   
Assuming pressure support, the total mass of the central concentration is $>10^{10.2}\,M_\odot$, while a lower limit to the dynamical mass in the case of full rotational support is $10^{10.4}\,M_\odot$. If the central concentration is a stable structure, it has to contain some amount of unseen matter, but potentially less than is observed for a typical galaxy. It is, however, not clear 
whether the structure has any gravitationally stable concentrations.  
We report a faint UV--optical--infrared source in proximity to one of the smaller concentrations in the gas complex, leading to a possible stellar association.
The system nature and origins is enigmatic, potentially being the result of an interaction with or within the galaxy group it appears to be associated with.
\end{abstract}

   \keywords{IGM --
                ISM --
                galaxies
               }
%

${ \_\, \,}$

\section{Introduction}\label{sec:introduction}
Some of the earliest radio observations already proved the existence of extragalactic neutral hydrogen in the form of intragroup gas or streams in nearby galaxy pairs or groups
\citep{roberts_neutral_1968,weliachew_neutral_1969,roberts_gaseous_1972, shostak_aperture_1974,van_der_hulst_kinematics_1979,van_der_hulst_structure_1979,appleton_neutral_1981}, the earliest being the Magellanic stream
\citep{hindman_low_1963,hindman_low_1963-1,roberts_neutral_1968,wannier_unusual_1972,mathewson_magellanic_1974,mathewson_intergalactic_1975,haynes_are_1979}. Such gaseous streams were identified as tidal features \citep{toomre_galactic_1972,davies_tidal_1977,haynes_detailed_1979,sancisi_neutral_1984,yun_high-resolution_1994,hibbard_dynamical_1995,barnes_transformations_1996, putman_tidal_1998}. They are not self gravitating or otherwise dynamically stable, and cannot exist in isolation.

In addition, it soon became clear that intergalactic \HI\ at the current level of column density sensitivity is generally a rarity. In a blind survey with the NRAO 91cm transit telescope, \citet{shostak_neutral_1977} were able to identify one local cloud with negative velocity, likely associated with the Magellanic Stream. 
\citet{lo_search_1979}, in an \HI\ survey of three galaxy groups, were not able to detect isolated clouds, but identified (4) new dwarf systems instead. 

Generally, despite counter-indications \citep{zwaan_targeted_2001}, it could be confirmed that galaxy groups, with a higher likelihood of galaxy encounters and hence a higher frequency of tidal interactions, harbour extraplanar or extragalactic gas more frequently \citep{hart_neutral_1980,sancisi_neutral_1984,appleton_neutral_1981,haynes_neutral_1981}, in particular in compact groups \citep{allen_low_1980,verdes-montenegro_where_2001,borthakur_detection_2010, 2017HessCluver}. 
Despite many galaxies being disturbed in their outer \HI\ disks \citep{huchtmeier_extended_1982,briggs_outlying_1982}, which extend well beyond their optical counterparts, no evidence was found that \HI\ disks of galaxies often connect to a low-column density (extragalactic) regime other than tidal debris \citep{briggs_first_1980, haynes_neutral_1981,haynes_influence_1984}. 
Blind \citep{lo_search_1979,krumm_neutral_1984, henning_study_1992,henning_study_1995, schneider_deep_1998,
hoffman_search_1992,mcmahon_h_1992,li_neutral_1994, sorar_new_1994,briggs_driftscan_1997, rosenberg_arecibo_2000, koribalski_atca_2003,kilborn_wide-field_2005} and pointed \citep{fisher_neutral_1981,fisher_upper_1981} \HI\ surveys were analysed for the abundance of isolated clouds in various environments, bearing a close to vanishing abundance of dark clouds at higher masses ($M_{\HI}\,>\,10^9\,M_\odot$).
\citet{zwaan_h_1997} were not able to detect massive, dark, isolated \HI\ clouds in the Arecibo \HI\ strip survey, and even HIPASS \citep{barnes_hi_2001} was not able to clearly detect any optically dark \HI\ clouds \citep{banks_new_1999,doyle_hipass_2005,karachentsev_optical_2008,wong_noircat_2009} not identifiable as tidal debris \citep{kilborn_extragalactic_2000,ryder_hipass_2001}. 

At the low-mass end, \citet{taylor_h_1995} searched for the existence of intergalactic clouds in the environment of blue compact dwarf galaxies, potentially triggering their starburst activity, without success \citep{taylor_h_1996}. Other studies found that there is even not much room for a gas-rich low-surface brightnes dwarf galaxy population \citep{briggs_space_1990,weinberg_population_1991}. While some surveys targeting the galaxy cluster environment seemed initially to be unsuccessful in detecting dark clouds \citep{dickey_vla_1997,barnes_hi_1997,van_driel_non-confirmation_2003,pisano_searching_2004}, the occasional lower-mass clouds could be found \citep{putman_fcc_1998}. It has since become well-established that the cluster environment, in which several mechanisms act to remove gas from galaxies and the paths connecting interaction partners are larger, is the preferred environment to find isolated \HI\ clouds \citep{hoffman_dark_1999,bravo-alfaro_vla_2000, bravo-alfaro_vla_2001}. Under which circumstances those clouds can exist as long-lived individual entities or whether they are tidal features (or both in the case of tidal dwarf galaxies) is an ongoing matter of debate.
The quest to determine the abundance of dark intergalactic clouds nevertheless continues to trigger new studies, as their existence is cosmologically relevant.

Firstly, the existence of dark matter substructure with masses below those of observed dwarf galaxies is a postulate of the standard galaxy formation theory \citep{kauffmann_formation_1993, klypin_where_1999,moore_dark_1999}. It appears therefore conceivable that dark galaxies exist, which, below a certain mass threshold cease to form stars completely. Dark \HI\ clouds could, in principle, contain gas that is gravitationally bound to a dark matter halo, without the ability to form stars \citep{verde_abundance_2002,taylor_star_2005}. Hence, the absence of low-mass galaxies or dark galaxies supports the conclusion that regulatory mechanisms, i.e. a photoionising background and supernova feedback deplete gas in the lowest-mass halos before a substantial amount of stars can be formed \citep{larson_effects_1974, rees_lyman_1986, quinn_photoionization_1996,barkana_photoevaporation_1999, efstathiou_model_2000,gnedin_fossils_2006}. 

With these regulatory mechanisms becoming a necessity to explain various luminosity and mass functions, the existence of dark galaxies would in turn be a problem for the standard theory. In fact, claims for the existence of massive dark galaxies or the detection of dark galaxy candidates \citep{davies_multibeam_2004,minchin_dark_2005,minchin_21_2007,walter_vla_2005, oosterloo_is_2013} triggered a substantial scientific discussion, mainly focusing on the question whether the velocity structure of tidal debris can, in projection, mimic a cold gas disk in stable rotation about a dark matter halo \citep{bekki_massive_2005,bekki_dark_2005,vollmer_ngc_2005,duc_tidal_2008,taylor_arecibo_2013,taylor_kinematic_2017,taylor_simulating_2018}, rendering the interpretation of a dark cloud as a dark galaxy unlikely. 

Secondly, the abundance and mass distribution of neutral extragalactic gas in various environments and its stability against the various mechanisms of gas depletion can be used to gauge galaxy formation scenarios.
While with increasing sensitivity the number of serendipitous detections of dark clouds or condensations in tidal debris increased \citep{hibbard_hi_2001,hibbard_high-resolution_2001, koribalski_neutral_2004,McKay04,koribalski_hi_2004,koribalski_neutral_2005, kilborn_gaseous_2006, english_vela_2010}, only recently have a substantial number of dark clouds been found in blind \HI\ surveys \citep[cf.][]{Wong21}.

Most notably, using the 40\% release data of the blind ALFALFA survey  \citep{giovanelli_arecibo_2005}, \citet[e.g.,][]{cannon_alfalfa_2015} estimated that less than 1.5\% were not associated with stars, and 25\% of those could not immediately be identified as tidal features. This triggered a number of follow-up studies of the most intriguing cases \citep{cannon_alfalfa_2015, janowiecki_almost_2015, ball_enigmatic_2018, Leisman21}, among others unveiling the characteristics of the extreme high-\HI\ mass LSB  object AGC~229101, with $M_{\textrm{H}\textsc{I}}/M_{\star}\approx 70$ and $M_{\textrm{H}\textsc{I}}\,=\,10^{9.3}M_\odot$. None of the sources has been found to lack any stellar counterpart, while \citet{kent_optically_2007, kent_clouds_2010} confirmed the detection of two formerly unknown dark clouds with $M_{\textrm{H}\textsc{I}} \le 10^{8.5}\,M_\odot$ in the vicinity of the Virgo cluster.
The second survey turning out a substantial number of dark cloud candidates is AGES \citep{auld_arecibo_2006}. \citet{taylor_arecibo_2012} found 7 ($M_{\textrm{H}\textsc{I}} \le 10^{7}\,M_\odot$) dark clouds in the surroundings of the Virgo Cluster. 
\citet{taylor_attack_2016} provide a comprehensive list of known dark clouds including AGES detections. Remarkably, dedicated surveys of two rich groups, CVn \citep{kovac_blind_2009}, and UMa \citep{lang_first_2003,wolfinger_blind_2013}, 
detected two small isolated clouds with $M_{\textrm{H}\textsc{I}} \le 10^{7.4}\,M_\odot$.

In summary, massive ($M_{\HI} > 10^{9}\,M_\odot$) optically dark neutral hydrogen clouds (or, rather, candidates for dark clouds) are so rare that they are the subject of individual studies. Most notable, beneath AGC 229101 \citep{Leisman21}, are the Leo\,Ring \citep{schneider_discovery_1983,schneider_neutral_1985,schneider_high-resolution_1986,schneider_neutral_1989} and the double source HI\,1225+01 \citep{giovanelli_protogalaxy_1989}. 

In this paper, we report the discovery of a massive \HI\ cloud that does not appear to have any star formation history, located in a filamentary large scale structure, and in close proximity to a galaxy group.  In the following we describe the radio observations, source extraction, and comparison with deep multi-wavelength imaging to discern the nature of the object.
Throughout this study, if not stated otherwise, systemic velocities are quoted in the optical convention $v\,=\,cz$, while line widths are given in the local reference frames. We assume a standard $\Lambda$CDM cosmology with $\Omega_\Lambda\,=\,0.73$, $\Omega_\mathrm{M}\,=\,0.27$, and $H_0\,=\,70\,\mathrm{km}\,\mathrm{s}^{-1}$.

%

\section{Observations and data reduction}\label{sec:datared}
\begin{table}
    \centering
    \begin{tabular}{@{} lll @{}}
       \multicolumn{2}{c}{Observational Parameters}\\
       \hline
       \hline
	Parameter &  \\
       \hline
    Target &  MeerHOGS\\
       	Observing dates                    & 17,24,26,31 May 2019\\
       	Number of pointings                & 25\\
       	Bandpass/flux calibrator           & PKS 1934-63\\
       	Gain calibrator                    & J2302-3718\\
       	Time spent on each pointing        & 30.4 min \\
       	Total observation time             & 4.5,4,4,4 h \\
        Available frequency range          & $900 - 1670\, \mathrm{MHz}$ \\
        Frequency range used               & 1319.8--1517.1\,MHz\\
       	Central frequency                  & $1416.8\,\mathrm{MHz}$\\
       	Spectral resolution                & 208.984 kHz \\
       	Available number of channels       & 4096\\
       	Number of channels used            & 959\\
       	Number of antennas                 & 58,58,64,58\\
        rms continuum, per pointing              & $18\,\mathrm{\mu Jy}\,\mathrm{beam}^{-1}$\\
        rms line mosaic                   & $0.19\,\mathrm{mJy}\,\mathrm{beam}^{-1}$\\
       	Mosaic spatial resolution (HPBW)   & \\
       	continuum & $13.\!\!\arcsec5 \,\times\, 13.\!\!\arcsec5$ \\
        line & $34.\!\!\arcsec4 \,\times\, 34.\!\!\arcsec4$ \\
 \hline
    \end{tabular}
     \caption{The details of the MeerKAT observations used in this study. See Fig.\ref{fig:mosaic_layout} for the combined footprint of the 25 pointings. Notice that for a mosaic the rms is variable. In this table we indicate a typical rms as relevant for this study.}
    \label{tab:obs_info}
\vspace{-10pt}
\end{table}

The MeerKAT Habitat of Galaxies Survey (MeerHOGS) is an initiative to exploit the sensitivity and 1$\degr$ field of view of the MeerKAT SKA Precursor to map local large scale structures. This study is based on the survey pilot, which targeted a $\sim$10\,deg$^2$ cosmic filament at redshift $0.025\,< z \, < 0.034$, visible in the redshift distributions of the 2dFGRS \citep{2dfGRS} and Galaxy and Mass Survey (GAMA) G23 surveys \citep{2011GAMA, 2015GAMA}. For details on the survey layout and the data reduction we refer to Appendix~\ref{appendix_layout}. Here, we provide a summary. 

The observations were centred on the highest concentration of galaxies in the filament which included a compact group identified using the 2MRS \citep{diaz-gimenez_compact_2012} at $z\approx 0.029 $. Within the GAMA survey, this 4-member compact group lies at the core of a $N=18$ group with a friends-of-friends redshift, $z_\mathrm{fof}=0.02891$, identified in the  GAMA Galaxy Group Catalog (G$^3$C) which employs an iterative friends-of-friends algorithm \citep{GAMAgroups}.  Henceforth we will refer to this group as `Group-83' to match the identification in the G$^3$C.

\begin{figure*}[ht]
\hspace{-20pt}
\includegraphics[width=1.05\textwidth]{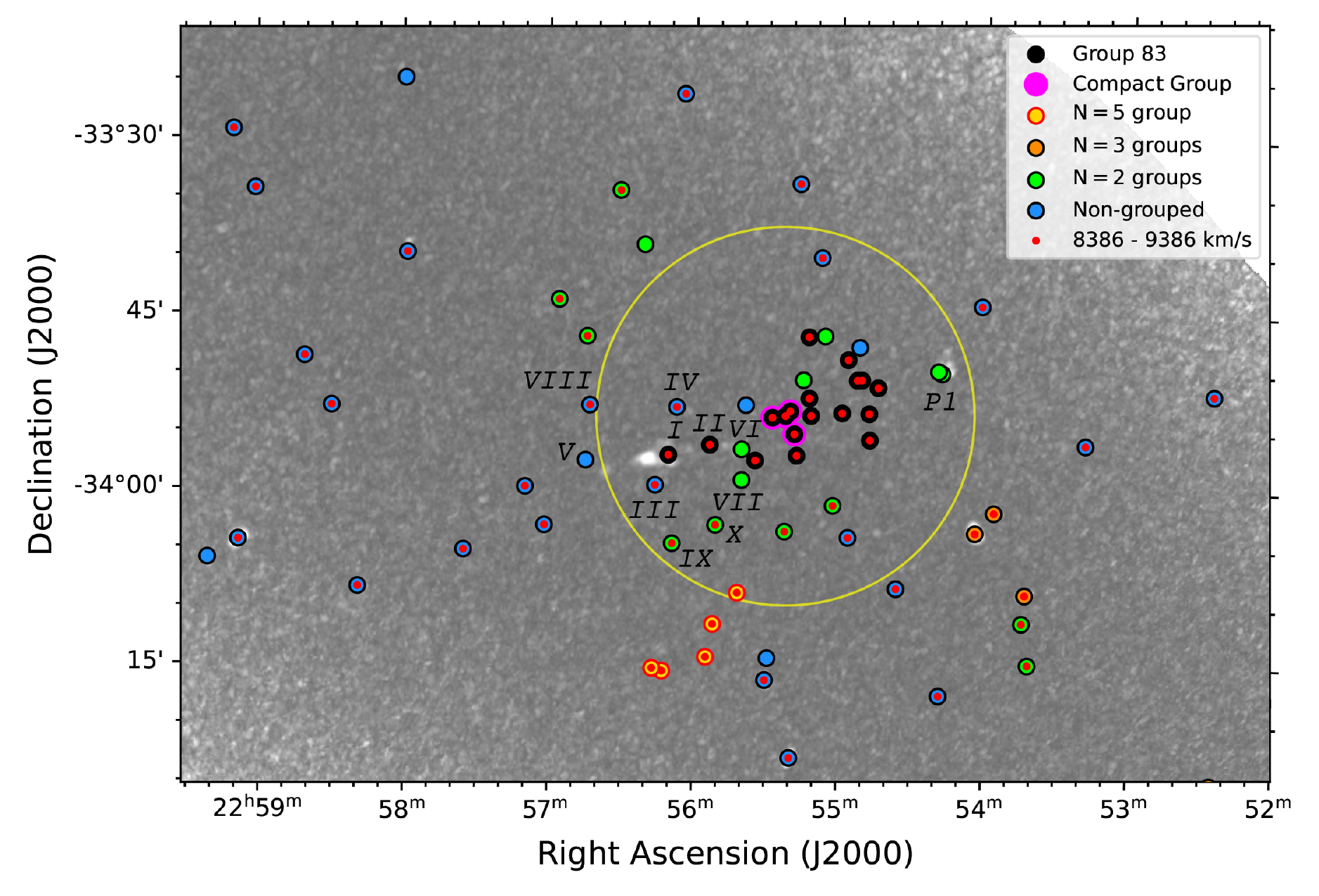}
\vspace{-25pt}
\caption{The immediate environment of the dark chain within the $0.025\,< z \, < 0.034$ redshift range, highlighting different groupings of objects in the filamentary structure.  The greyscale image is the \HI\ moment-0 map whose column density ranges between 1 to 25$\,\times\,$10$^{19}$\,atoms\,cm$^{-2}$, the dark cloud chain is the most massive gas complex in the region.
The black points denote GAMA Group 83, a $N=18$ group with $z_\mathrm{fof}\,=\,0.02891$ (8667 km\,s$^{-1}$), where the yellow circle (centred on the group) is 0.26\,degrees, $\sim$0.56\,Mpc radius, which corresponds to the R$_{200}$ of Group-83. Galaxies with velocities within $\pm$500\,km\,s$^{-1}$ of the dark chain (8886\,km\,s$^{-1}$). Labels I-X indicate the position of galaxies in the projected proximity of the dark \HI\ chain (9\arcmin $\sim$320 kpc) as discussed in Sect.~\ref{sec:results}.  See Table~\ref{tab:enviro_prop} for additional information.
}
\label{fig:meerhogs_massive_dark_cloud_enviro}
\end{figure*}

The filament was observed in 20 pointings arranged in a linear fashion along the filament, with 5 additional pointings in the vicinity of Group-83, in L-band with the MeerKAT telescope in May 2019. The usage of the available 4K correlator mode resulted in a frequency resolution of $209\,\mathrm{kHz}$ or 45.4\,km\,s$^{-1}$ in the restframe of an object at z\,$\sim$\,0.029. The integration time for each pointing was $\sim$30\,min. While the L-band frequency range of MeerKAT is $900 - 1670\, \mathrm{MHz}$, we used only the frequency range 1319.8--1517.1\,MHz for the data reduction with the 
{\sc CARACal} \citep{Jozsa2020} radio-interferometric data reduction pipeline. {\sc CARACal} performs automated standard calibration and imaging, including 
flagging, crosscalibration, self-calibration, deconvolution and mosaicing. For the latter, we additionally employed the {\sc equolver} package (see Appendix \ref{appendix_layout}) at the time of the data reduction, not yet implemented in {\sc CARACal}) convolving the images to a common resolution before mosaicing.  The source-finding software {\sc SoFiA} \citep{serra_sofia:_2015} 
was combined with a Savitzky-Golay \citep{savitzky_1964} 
filter to perform an additional continuum subtraction in the image domain, as well as to identify \HI\ sources in the data cube. Table~\ref{tab:obs_info} summarizes the observational parameters. For the channel maps used in this study we obtained an rms of $0.19\,\mathrm{mJy}\,\mathrm{beam}^{-1}$ at a spatial resolution of $34.\!\!\arcsec4 \,\times\, 34.\!\!\arcsec4$.

%

\section{Results}\label{sec:results}

%

\subsection{Galaxies in the Vicinity of the Dark Cloud Chain}
\begin{table*}[!ht]
    \centering
    \begin{tabular}{llccccc}
       \multicolumn{4}{l}{Investigating Possible Associations of the Dark Chain}\\
       \hline
       \hline
Identifier &	CATAID & RA & Dec & $z_\textrm{helio}$ & $cz$  \\
--  &     -- &   deg & deg & -- & km\,s$^{-1}$ \\
\hline
\hline
Group-83 Galaxies\\
\hline
Galaxy \textit{I}  & 5256498 & 344.04430   &  -33.95156  & 0.02805  & 8409 $\pm$ 85  \\
Galaxy \textit{II}  &  5256643 &  343.97295 & -33.93668 & 0.02860    & 8574 $\pm$ 25  \\
\hline
Ungrouped Galaxies\\
\hline
Galaxy \textit{III}  &  5256376 &  344.06646  & -33.99435 & 0.02920   &  8754 $\pm$ 85 \\
Galaxy \textit{IV} &  ---$\dagger$ & 344.03040  & -33.88321 & 0.02806   &  8412 $\pm$ 85 \\
Galaxy \textit{V} &  5256491 & 344.18659  &   -33.95962 & 0.03274   &   9815 $\pm$ 25  \\
Galaxy \textit{VIII} &  5256924 & 344.17966  & -33.88105 & 0.02820   &  8454 $\pm$ 85 \\
\hline
Pairs\\
\hline
Galaxy \textit{VI} & 5256517 &  343.91816  &   -33.94274 & 0.02723   &   8163 $\pm$ 85 \\
Galaxy \textit{VII} & 5256426 &  343.91790 & -33.98653  & 0.02730     & 8184 $\pm$  35  \\
\hline
Galaxy \textit{IX} & 5255926  & 344.03695  & -34.07758 & 0.02814   &    8436 $\pm$  85  \\
Galaxy \textit{X} & 5256135 & 343.96247  &  -34.05080 & 0.02868    &   8598 $\pm$ 25  \\
\hline
Pair 1  & 5241095  & 343.58043   &  -33.82934  & 0.02742     &   8221 $\pm$ 25 \\
        & 5240983  & 343.57447   &  -33.83238  & 0.02733    &   8194 $\pm$ 24 \\
\hline
\hline

\end{tabular}
\caption{Galaxies within projected proximity of the dark \HI\ chain (9\arcmin $\sim$310 kpc at the distance of the central concentration of the \HI\ cloud chain).  The CATAID corresponds to the GAMA identification and V$_{\textrm{optical}}$ corresponds to the heliocentric velocity.\\
$\dagger$ This galaxy does not have a CATAID as it was not included in the GAMA main survey sample selection \citep{Baldry2010}; the redshift is from the 2dFGRS \citep{2dfGRS}
}
\label{tab:enviro_prop}
\end{table*}

Initial \HI\ source finding was carried out with SoFiA (see Sect.~\ref{sect_source_finding}).
We identified 62 \HI\ sources (of 196 total detections in our \HI\ cube) whose systemic velocities are in the redshift range of the filament:  0.025 $< z <$ 0.034 (7500--10\,201\,km\,s$^{-1}$). 
We compared the resultant source positions and velocities with the GAMA G23 redshift catalogue using a 10\arcsec\ ($\sim$ 6 kpc at z$=0.0295$) search radius and $\pm$180\,km\,s$^{-1}$ cylinder, finding 46 associations with known galaxies. These associations include galaxies from the 2dFGRS \citep{2dfGRS} that were not included as part of the GAMA science sample. Only galaxies with redshift quality flags NQ$\geq3$, i.e science quality, are considered.

During the course of identifying associations, however, we noted a number of \HI\ sources without optical counterparts. Constructing a moment-0 map from the \HI\ cube in the velocity range of the filament, we then compared it to deep optical imaging from the KiDS survey \citep{Kui19}, sensitive 3.4\,$\mathrm{\mu}$m mid-IR imaging from WISE \citep{Wri10} and FUV/NUV imaging from GALEX \citep{CMar05}. We discovered a relatively massive \HI\ source (or complex of sources) that did not have a definitive counterpart in the UV, optical, or infrared imaging, nor in the radio continuum image produced from these MeerKAT observations (Yao et al. ApJ submitted).  We dub this complex the ``dark cloud chain" or ``dark cloud complex" throughout this paper.

In the rest of this section we lay out the nature of this dark cloud complex as follows.  
We first consider the larger environment to put the cloud complex into context, particularly with respect to the neighbouring galaxy groups.
We then demonstrate the robustness of the \HI\ cloud complex, including after-the-fact confirmation with HIPASS data. Finally, we limit the possibility of stellar counterparts using NUV, optical, and IR imaging.

MeerHOGS was designed to target a foreground large scale filament at $z\sim  0.03$, in which the dark cloud chain is centrally located.  
We consider the environment of the dark chain to ascertain its possible origin scenarios, focusing only on the redshift range $0.025 < z < 0.034$. Within the GAMA G23 region we benefit from high redshift completeness,  98\% for 19.2\,mag in \textit{r}-band \citep{2015GAMA}, and therefore superior environment measures; for instance the GAMA Galaxy Group Catalog (G$^3$C) which employs an iterative friends-of-friends algorithm \citep{GAMAgroups}.

Figure~\ref{fig:meerhogs_massive_dark_cloud_enviro} shows a \HI\ moment-0 map of the central region of the MeerHOGS field where our \HI\ observations are the most sensitive.
The dark cloud chain is located within or very close to GAMA Group-83 (black points), a $N=18$ galaxy group with a `friends-of-friends' redshift of 0.02891 (8667\, km\,s$^{-1}$). The originally targeted compact group is at its core (magenta points in Fig.~\ref{fig:meerhogs_massive_dark_cloud_enviro}), with the BCG (Brightest Cluster Galaxy) of the group as one of the members of the compact group (which includes the galaxy, IC\,5262). 
Other galaxies within the G$^3$C in this redshift range are demarcated according to their group size, (green, orange, and yellow) with non-group galaxies shown as blue points. Labels I-X indicate the position of galaxies in the projected proximity of the dark \HI\ chain (9\arcmin $\sim$320 kpc); we provide additional information for these select galaxies in Table~\ref{tab:enviro_prop}, including the GAMA identification and coordinate position, optical redshift, and velocity. Velocity errors from the GAMA-AAOmega spectra are individually calibrated for each spectrum using AUTOZ \citep{Baldry14}, whereas redshifts observed as part of 2dFGRS subsumed into the GAMA survey \citep{Baldry2010} are assigned an error of 85\,km\,s$^{-1}$ \citep[see][]{2dfGRS}.

The gas complex ranges in column density up to $25\,\times\,$10$^{19}$\,atoms\,cm$^{-2}$ and has a systemic velocity of $\sim$ 8886\,km\,s$^{-1}$ (z\,$=$\,0.0296); galaxies within 500\,km/s of the complex are marked with a red dot in Fig.~\ref{fig:meerhogs_massive_dark_cloud_enviro}. 
The dark chain lies at the eastern edge of the large group, Group-83, notably close to member \textit{I} (z\,$=$\,0.02805) and member \textit{II} (z\,$=$\,0.02860); see Table~\ref{tab:enviro_prop}. Further to the group members are the non-grouped galaxies \textit{III, IV} and \textit{V} with redshifts of 0.02920, 0.02806 and 0.03274, respectively. We note that no member of Group-83, nor galaxy \textit{III} or \textit{IV} have an \HI\ detection. Galaxy \textit{V} is therefore the exception, but with a redshift well distant of the dark cloud. The only other notable detection of \HI\ in this region is the interacting pair of galaxies (P1 in Fig.~\ref{fig:meerhogs_massive_dark_cloud_enviro}; Pair 1 in Table ~\ref{tab:enviro_prop}) on the opposite side of Group-83 to the dark chain. However, these galaxies lie in front of the dark chain, offset by $\sim$600 km \,s$^{-1}$. 

\begin{figure*}
\includegraphics[width=0.99\textwidth]{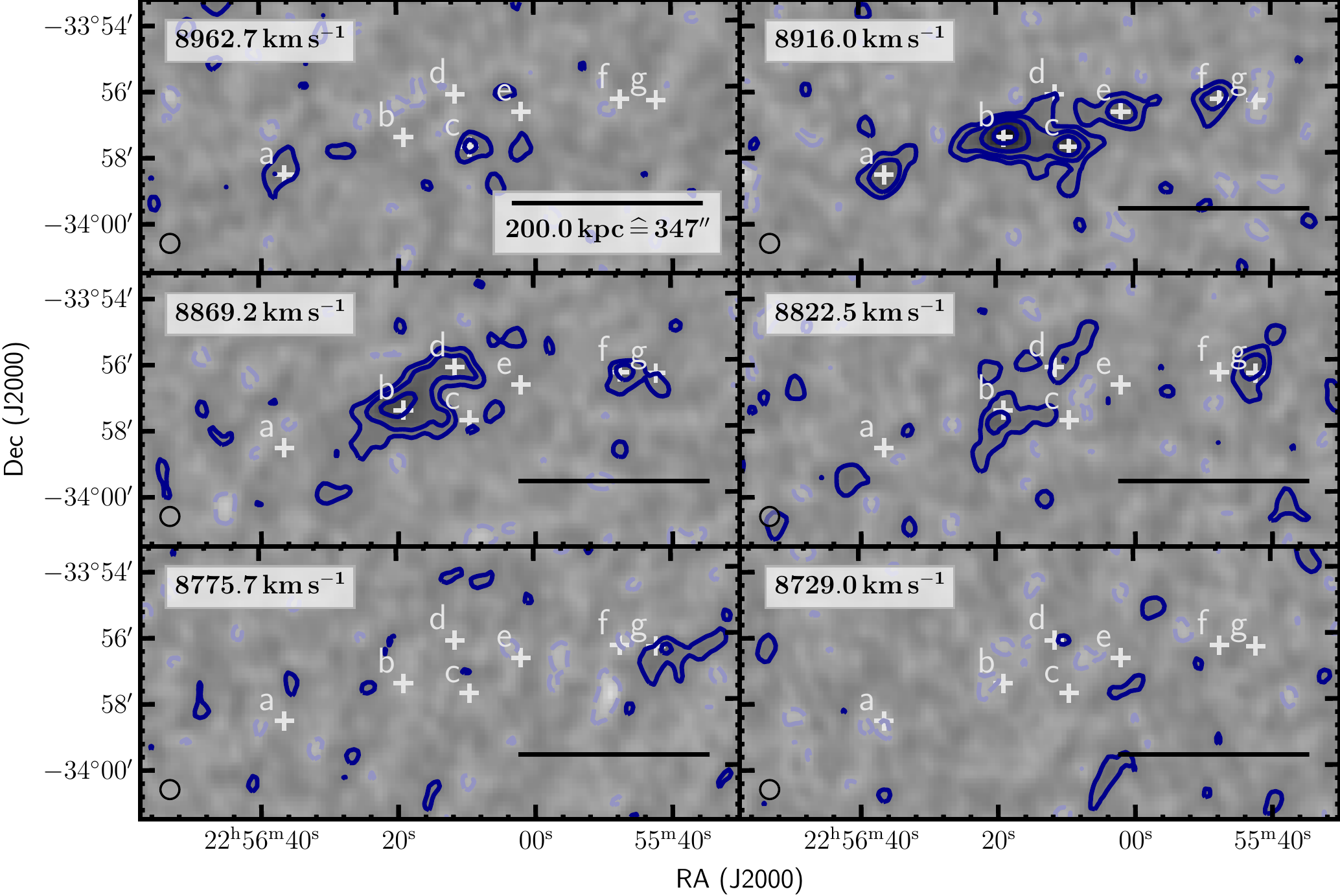}
\vspace{-5pt}
\caption{Observed \HI\ data cube in the region and velocity range of the dark cloud chain. Contours denote the -0.38, 0.38, 0.76, 1.52, and 3.04 $\mathrm{mJy}\,\mathrm{beam}^{-1}$ levels (-2, 2, 4, 8, 16 $\sigma_\mathrm{rms}$), negative contours are dashed. White crosses mark \HI\ detections (labeled a-g; see Table \ref{tab:HIprop}). The black circle in the lower left of each panel indicates the spatial resolution (HPBW). The scale bar indicates 200\,kpc.
The massive central concentration (Source b) has a peak intensity in the second velocity channel (8916\,km\,s$^{-1}$).}
\label{fig:datacube}
\end{figure*}

We note that the R$_{100}$ (i.e. the group radius defined by the most distant member from the iterative centre determined for the group) for Group-83 is 0.36\,Mpc, and has a derived velocity dispersion of $228.6\,\mathrm{km\,s}^{-1}$ (similar to the cloud velocity width) and dynamical mass of $10^{13.5}\,M_\odot$.
Further, assuming an isothermal distribution, the $R_\mathrm{200}$ -- which follows from the relation $\frac{\sqrt{3}\,\sigma_\mathrm{gr}}{10\,H(z)}$ -- has a value of 0.56\,Mpc (shown as the yellow circle in Fig. \ref{fig:meerhogs_massive_dark_cloud_enviro}). Further, we note the velocity difference between the central gas concentration and the Group-83 center is $\sim$200\,km\,s$^{-1}$, which is smaller than the group dispersion. Consequently, the dark cloud chain is well within the gravitational influence of the Group, and likely the compact core at the center. 

%

\subsection{H {\sc i} Properties of the dark cloud chain}
\begin{table*}
\hspace{-65pt}
    \begin{tabular}{ccccccccccccc}
       \multicolumn{13}{c}{\HI\ Detections in the Cloud Chain}\\
       \hline
       \hline
src & designation & RA DEC & $V_\mathrm{radio}$ & $V_\mathrm{opt}$ & $w_\mathrm{x}$ & $w_\mathrm{y}$ & $w_\mathrm{z}$ & $w_{20}$ & $I_\mathrm{peak}$ & $F_\mathrm{tot}$ & $D_\mathrm{L}$ & Log$M_\mathrm{\HIe}$\\
	-- &     --     & deg  & km\,s$^{-1}$ & km\,s$^{-1}$ & \arcsec & \arcsec & km\,s$^{-1}$ & km\,s$^{-1}$ & 10$^{19}$\,cm$^{-2}$ 
	              & Jy\,km\,s$^{-1}$ & Mpc & M$_\odot$ \\
\hline
\hline
(a) & J22563653-3358328 & 344.15222 -33.97579 & 8673 & 8931 & 80& 80 &136 & 106&   8.77  &   0.200 &126.6 & 8.88 \\    
(b) & J22561910-3357270 & 344.07962 -33.95752 & 8630 & 8886 &200&128 &182 & 139&  26.26  &   1.353 &125.9 & 9.70 \\    
(c) & J22560952-3357456 & 344.03967 -33.96267 & 8657 & 8914 &104& 88 &182 & 105&  16.05  &   0.374 &126.4 & 9.15 \\      
(d) & J22561155-3356098 & 344.04813 -33.93608 & 8604 & 8858 &104& 80 &182 & 132&  10.98  &   0.357 &125.5 & 9.12 \\        
(e) & J22560193-3356425 & 344.00806 -33.94515 & 8651 & 8908 & 88& 72 &136 & 86 &   7.32  &   0.181 &126.3 & 8.83 \\        
(f) & J22554753-3356212 & 343.94806 -33.93924 & 8639 & 8895 & 72& 56 &182 & 110&   8.85  &   0.151 &126.1 & 8.75 \\      
(g) & J22554226-3356236 & 343.92612 -33.93991 & 8565 & 8817 & 64& 64 &182 & 135&   8.18  &   0.189 &124.9 & 8.84 \\
\end{tabular}
\vspace{-5pt}
\caption{\HI\ Measurements and Source Properties of the dark cloud chain.  See Fig.~\ref{fig:4plot} for Source (a)-(g) identifications.  The formal ``designation" should have the prefix ``MHGS" to signify a MeerHOGS source.
The widths represent the mask limits (x, y, z respectively) of the source measured from the \HI\ cube; see more details of the mask in the Appendix~\ref{appendix_b}.  The noise level (per voxel) of the \HI\ distribution in vicinity of the dark cloud chain is $\sim 0.9\,\times\, 10^{19}\,\mathrm{atoms}\,\mathrm{cm}^{-2}$, and the uncertainties for the integrated fluxes are $<$10\%, although the blending systematics between sources will dominate the accuracy of the fluxes and the derived \HI\ mass. No attempt has been made to correct $w_\mathrm{z}$ or $w_{20}$ for instrumental broadening. 
}
\label{tab:HIprop}
\vspace{-5pt}
\end{table*}

First shown at the center of the \HI\ map
(Fig.~\ref{fig:meerhogs_massive_dark_cloud_enviro}), the \HI\ dark cloud complex extends approximately 10\arcmin\ from east to west and across only a few ($\sim$3 to 4) channels in velocity, or roughly 
$<$\,136--182\,km\,s$^{-1}$ in width, for any given region of the cloud complex. This is also shown in Figure~\ref{fig:datacube}, a cutout of our data cube in the region and velocity range of the cloud complex.  The channel maps reveal very little kinematic variation across the cloud
from east-to-west. 

We will now focus on characterizing the gas cloud complex using an integrated map that delineates the diffuse from clumpy components.

For a range in  velocity that takes into account this lateral east-west geometry of the cloud relative to the plane-of-the-sky, 8682 to 9057\,km\,s$^{-1}$, 
we construct a detailed view of the cloud complex using a moment-0 mapping technique in which we integrate 3 channels along the line of sight where the emission is peaking. 
In this way we maximize the emission S/N for each spatial pixel of the \HI\ map without the use of a mask. Under the usual assumption that the observed \HI\ is optically thin, the moment-0 map is then converted into a column-density map in units of $\mathrm{atoms}\,\mathrm{cm}^{-2}$.

\begin{figure*}
\includegraphics[width=1\textwidth]{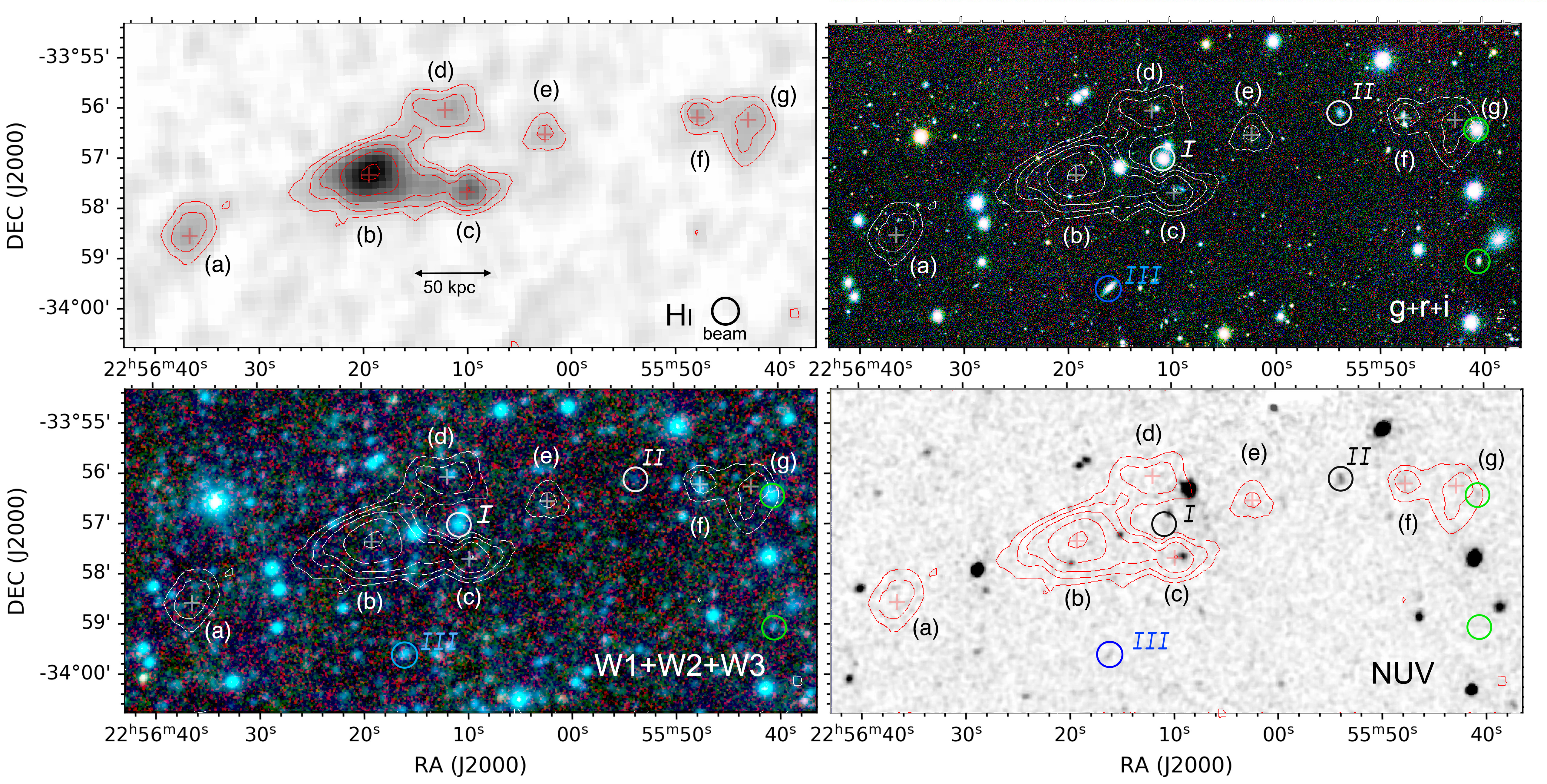}
\vspace{-15pt}
\caption{Dark cloud chain as seen across the electromagnetic spectrum.  The four panels show the  \HI\ column density map (upper left; 8817--8931\,km\,s$^{-1}$ with rms noise $\sim$\,0.9$\,\times\,$10$^{19}$\,atoms\,cm$^{-2}$),  
KiDS \textit{g,r,i}-bands (upper right), WISE  3.4-12\,$\mathrm{\mu}$m bands (lower left) and GALEX NUV (lower right). Contours of the \HI\ column density are overlaid on all four maps, with values of 3, 5, 8, 14, and 23$\,\times\,$10$^{19}$\,atoms\,cm$^{-2}$. Red/white crosses demark \HI\ detections (labeled a-g; see Table \ref{tab:HIprop}), and the small circles denote filament galaxies as marked in  Fig.~\ref{fig:meerhogs_massive_dark_cloud_enviro} (blue is non-group; black/white is Group; green are galaxy pairs). The closest Group-83 galaxy to the \HI\ is labeled \textit{I} (z$=$0.02805/8415\,km\,s$^{-1}$), just west of the massive gas complex.  Other nearby group galaxies are labels II (z$=$0.0286/8580\,km\,s$^{-1}$) and III (z$=$0.0292/8760\,km\,s$^{-1}$).
The central \HI\ source (Source b) is not detected in any band.
}
\label{fig:4plot}
\end{figure*}

The resulting detailed multi-wavelength view of the immediate complex is shown in Fig.~\ref{fig:4plot}, featuring  in the first panel the \HI\ moment-0 map overlaid with red contours of gas column density (3, 5, 8, 14 and 23$\,\times\,$10$^{19}\,\mathrm{atoms}\,\mathrm{cm}^{-2}$). The rms noise is about 0.9$\,\times\,$10$^{19}\,\mathrm{atoms}\,\mathrm{cm}^{-2}$, and hence the lowest contour is approximately at the 3$\,\sigma$ level.
We identify 
 at least seven local maxima, labeled a-g in Figs.~\ref{fig:datacube} and \ref{fig:4plot}.  The single constituents of the complex have a similar 
shallow kinematic depth of 3 to 4 channels, or $<$\,182\,km\,s$^{-1}$,
partly showing spurious kinematic signatures indicating systematic motion in parts of the whole complex (most notably between Sources f and g), although a general tendency is hard to discern. More clearly, the whole complex, appearing in five channels ($\sim$\,227\,km\,s$^{-1}$) shows an east-west velocity gradient (refer also to Fig.~\ref{fig:datacube});
note that the complex appears in 5 channels, while the depth for any given location in the cloud is 3 to 4 channels in width), again rendering the detection a real feature, as this would be hard to mimic by measurement errors or data reduction artifacts.

\begin{figure*}
\includegraphics[width=1\textwidth]{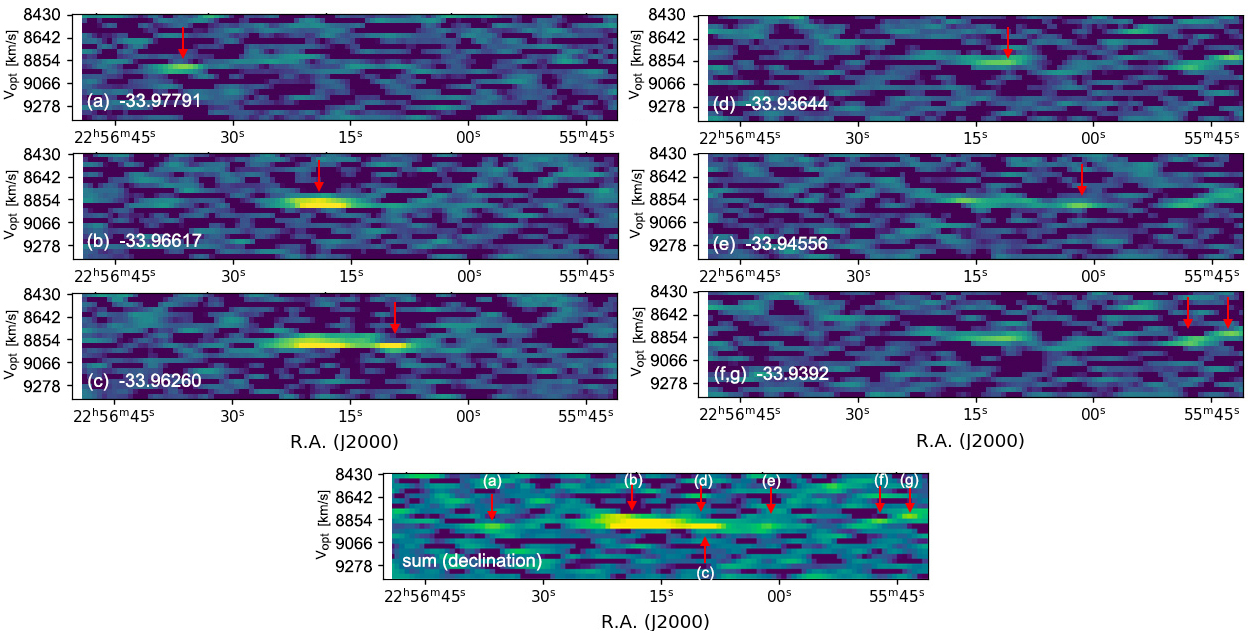}
\vspace{-20pt}
\caption{Position-velocity diagram for the dark cloud chain.  The top six panels have specific Declinations (Deg J2000) to highlight the \HI\ emission maxima: (a) to (g), which run from East to West. The bottom panel is the sum across the declination channels -33.990 to -33.936 [deg]. 
The red arrows indicate where the maxima are located in the RA-Velocity plane.  Note the filamentary structure along the East-West direction, and the lack of kinematic variation across the cloud complex.
}
\label{fig:PV}
\vspace{-10pt}
\end{figure*}

Supplemental to the channel maps and the multiwavelength maps, we present a RA-Velocity diagram that again reveals the dark cloud chain emission maxima in addition to its kinematical structure.  The diagram, shown in Figure~\ref{fig:PV}, has panels (slices) of Declination, chosen to highlight each individual maxima, from (a) to (g); see also the moment-0 map in Figure~\ref{fig:4plot} and the 3D version shown below in Figure~\ref{fig:idavie-horiz}.  The diagram clearly shows the resolved, filamentary or chain-like morphology of the gas, extending from east-to-west.  It is also clear that there is a lack of kinematic structure (it is relatively flat in velocity, $\sim 110\,\mathrm{km}\,\mathrm{s}^{-1}$ ($108\,\mathrm{km}\,\mathrm{s}^{-1}$ from source (a) to source (g), see also Table~\ref{tab:HIprop}) and shows little gradient from one declination slice to the next (best seen in Fig.~\ref{fig:idavie-horiz}).

Focusing on the first panel showing the \HI\ distribution (Fig.~\ref{fig:4plot}a)
the brightest and largest source is Source (b), which is resolved, while the other sources may be unresolved, but also embedded within diffuse emission.  
SoFiA was unable to automatically deblend this complex morphology. Consequently, we used a new visualisation tool to identify and divide the local maxima into seven discrete sources.  Developed specifically for spectral-imaging data interaction, and notably with \HI\ cubes, the Virtual Reality (VR) based software suite called \IDAVIE\ \footnote{\url{https://idavie.readthedocs.io/en/latest/}} \citep{Jar21} was deployed to separate the seven sources using mask-editing functionality in which the VR user is able to create (or modify) a mask that specifies which \HI\ intensity voxels (3D pixels) belong to which sources.  More details of this mask editing procedure and the resulting mask that was used to extract the source characterization is presented in Appendix~\ref{appendix_b}.  

Source extraction was then carried out using this mask created for the cloud complex.  Basic source properties are given in 
Table \ref{tab:HIprop}, including coordinates, spatial sizes, kinematics, column density and finally, integrated flux and corresponding \HI\ mass (see caption for details).  
Here the mass is derived from the integrated flux using the standard equation (assumes optically thin emission):  $\frac{M_\mathrm{\HIe}}{M_\odot} = 2.356 \,\times\, 10^5 \frac{F_\mathrm{tot}}{\mathrm{Jy}\,\mathrm{km}\,\mathrm{s}^{-1}}  \frac{D_\textrm{L}}{\mathrm{Mpc}}$, where the total flux, $F_\mathrm{tot}$, is integrated over all pixels and all channels (measuring the channel width in radio convention) and scaling for the beam factor, and the luminosity distance, $D_\textrm{L}$, in Mpc, using the optical redshift corrected to the CMB frame.
The total mass of the dark cloud chain is $10^{10.0}\,M_\odot$, with the masses of the concentrations ranging between $10^{8.75}\,M_\odot$ and $10^{9.7}\,M_\odot$.

The noise level (per voxel) of the \HI\ distribution in close vicinity to the dark cloud chain is $\sim$\,0.9$\,\times\,$10$^{19}$\,atoms\,cm$^{-2}$, and correspondingly the uncertainties for the integrated fluxes are well below 10\%.  However, since these sources are not well defined, but include both discrete and nebulous emission with connections between them, these source measurements are only approximate -- clearly there is overlap and blending systematics between sources that dominate the accuracy of the fluxes and the derived \HI\ mass.\\ 

\begin{figure}
\includegraphics[width=0.45\textwidth]{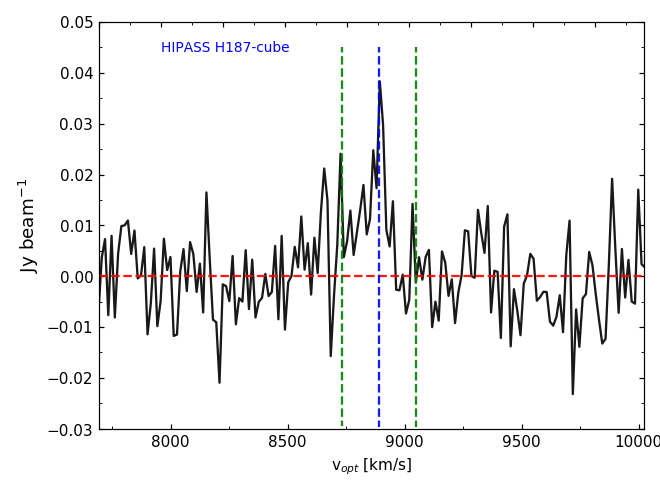}
\caption{\HI\ spectrum of the dark cloud chain as measured using the HIPASS cube H187.  The variation in the underlying continuum has been fit and removed.  The source is clearly detected at the spatial --- RA,DEC[deg J2000]\,=\,344.080, -33.958 --- and velocity coordinates of the massive \HI\ Source (b) (see blue vertical line at v$_{opt}$\,=\,8886\,km\,s$^{-1}$).  The green lines demark the velocity limits for the spectrum integral.
}
\label{fig:hipass2}
\vspace{-10pt}
\end{figure}
\underline{HIPASS:}  Such a massive \HI\ cloud may be marginally detected in the single-dish HIPASS survey.
We subsequently searched the HIPASS and HICAT catalogues \citep[see, for example,][]{barnes_hi_2001} for any counterparts.  There were none, and notably with few cataloged sources at such large systemic velocities.  
Searching through the HIPASS Public Data Release\footnote{https://www.atnf.csiro.au/research/multibeam/release/} spectra, we identified a possible detection in HIPASS cube ``H187" at the position of the dark cloud chain.
Prompted by this encouraging result, 
we obtained an improved cube used to make HICAT \citep{Meyer04}. We then extracted a spectrum centered on the spatial and velocity coordinates of Source (b).
We show the resulting spectrum in the velocity range of the cloud in Figure~\ref{fig:hipass2}. Details and a wide-band spectrum are given in Appendix~\ref{appendix_c}.  Here we summarize the results:
fitting and removing the underlying ripple continuum, we measure a  
peak flux, at  $\sim$8886\,km\,s$^{-1}$, to be 0.038\,Jy/beam with a S/N of $\sim$4.2.  Integrating the spectrum channels, across $\pm$150\,km\,s$^{-1}$ centered on the cloud systemic velocity, the integrated flux is 2.9$\pm$0.6\,Jy\,km\,s$^{-1}$; at the 5$\sigma$ level, a likely detection of the \HI\ gas cloud.  For the MeerKAT observations, we measure the integrated flux of Source (b) to be 1.4\,Jy\,km\,s$^{-1}$, and of the entire complex to be $\sim$2.8\,Jy\,km\,s$^{-1}$.  The single-dish results are therefore consistent with the aggregate interferometric measurements, given the large Parkes Telescope beam ($\sim$15.5\arcmin) and its sensitivity to low column density gas 
in the vicinity of the dark cloud chain. Moreover, the good agreement leaves very little room for an additional low-column-density component undetected by MeerKAT.\\

Next, we carried out a multi-wavelength analysis to interrogate the nature of the cloud complex.   In Figure~\ref{fig:4plot} (panels b, c, d, respectively) we compare the neutral gas distribution with deep \textit{g,r,i}-band imaging from KiDS-S \citep[\textit{g, r, i}-bands;][]{deJ17,Kui19}, WISE mid-infrared (W1, W2 and W3, covering 3.4--12\,$\mathrm{\mu}$m; \citet{Wri10}), and the ultra-violet GALEX (NUV; \citet{CMar05}),
where the red \HI\ contours of column density are repeated in all four panels for juxtaposition purposes. 

Comparing the \HI\ emission to the UV, optical, or IR imaging reveals that counterparts are not detected (with one possible exception, see below).  Most notably, the large central Source (b) has no counterpart whatsoever.  This is surprising given the measured gas mass -- $10{^{9.7}}$\,M$_\odot$, which is typical for a large, gas-rich galaxy -- and the 
depth of the imaging (details below), which suggests a very high gas mass-to-light ratio, and very little, if any, star formation history. The closest known galaxy is labeled \textit{I} in the diagram, a gas-poor passive galaxy located $>$70\,kpc from the cloud core (see Table~\ref{tab:enviro_prop}). The other bright sources in the field are foreground stars, most easily seen in the WISE 3-color image, Figure~\ref{fig:4plot}c  (stars very clearly appear as bright `blue' point sources).
Finally, we highlight Source (c), just to the west of the massive central cloud -- it is the second brightest source in the seven, and it may be associated with a faint optical galaxy (discussed below). 

The gas complex is located at the edge of the Group-83.
We find it intriguing that the dark cloud chain has a long filamentary appearance, stretching east to west, and pointing exactly toward the location of the core of the Group-83 (see Fig.~\ref{fig:meerhogs_massive_dark_cloud_enviro}).   
Given the relatively `flat' kinematics (Fig.~\ref{fig:datacube}) -- limited motion along the radial line of sight -- and the long chain extending over $\sim$400\,kpc (396\,kpc from Source a to Source g)
toward the group center
(Figs.~\ref{fig:meerhogs_massive_dark_cloud_enviro},\ref{fig:4plot}), it suggests a tantalizing dynamic link to the galaxy overdensity, taking into account the slight gradient along the filament, either lateral accretion toward it or expulsion away from it.   
However, neither the nature of the dark cloud, nor its physical connection to the group is certain.   
Next, we consider the  multi-wavelength information in quantitative detail.

\begin{figure*}
 \hspace{-10pt}
\includegraphics[width=1.02\textwidth]{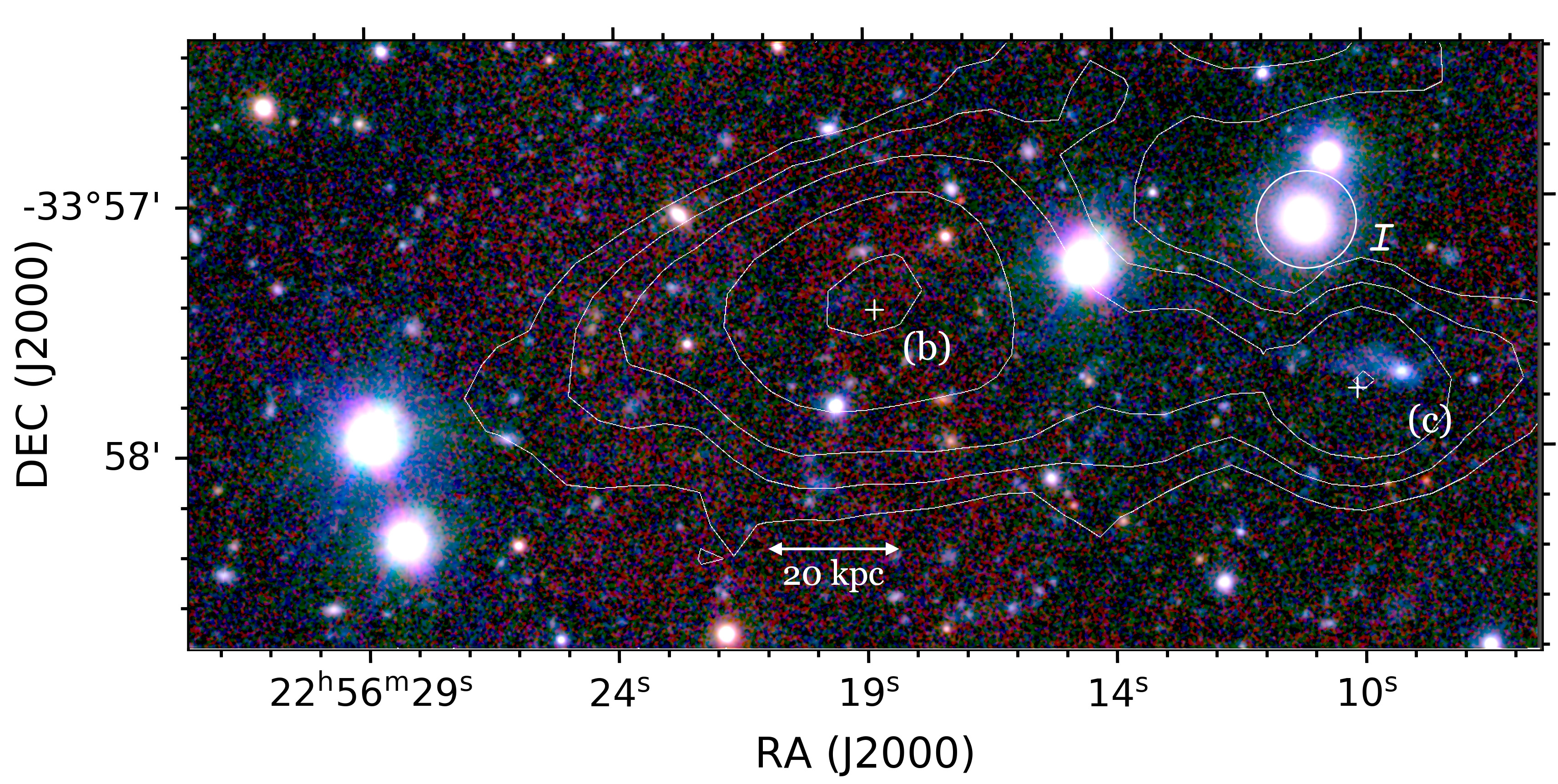}
\vspace{-15pt}
\caption{Central concentration of gas in 
the dark cloud chain shown in Fig.~\ref{fig:4plot}ab.  The image shows the 
 KiDS \textit{g,r,i}-bands (blue, green, red), overlaid with
 contours of the \HI\ column density: 3, 5., 8, 14 and 23$\,\times\,$10$^{19}$\,atoms\,cm$^{-2}$. White crosses demark \HI\ detections for Source (b) and Source (c).  
 The closest Group-83 galaxy to the Chain is labeled \textit{I}.
There are no clear optical associations with the densest gas in the entire region (Source b).   However for Source (c) we do detect a small `blue' dwarf galaxy; see Fig~\ref{fig:westsrc} for a detailed view of Source (c).
}
\label{fig:1plot}
\end{figure*}

%

\subsection{Massive Dark Cloud: Gas versus Star Formation History}

The chain of \HI\ ranges in peak column density from 7 to $25\,\times\,10^{19}\,\mathrm{atoms}\,\textrm{cm}^{-2}$, and in integrated masses from $10^{8.7}$ to $10{^{9.7}}\,M_\odot$; see Table \ref{tab:HIprop}.
For normal ranges of gas mass relative to optical luminosity (M/L), and column densities that indicate relatively clumpy gas,  these objects should be easily detected in the optical, assuming they have formed stars as galaxies.   
The optical \textit{g, r, i}-bands are sensitive to 
moderate-age (5\,Gyr) starlight and to young (10\,Myr) star formation (through the blue continuum and H$\alpha$+[NII] emission).  At 3.4\,$\mathrm{\mu}$m, the infrared map is sensitive to the old (11 to 13\,Gyr) stellar population, while the NUV map is sensitive to recent (100\,Myr) star formation; hence the set of maps cover most of the star formation history of a galaxy.
Figure~\ref{fig:4plot} reveals very little correspondence between the gas and extragalactic emission at any wavelength.

If we consider first the footprint of the dark cloud, denoted as (b) in Figure~\ref{fig:4plot}a, it extends over 50 kpc in diameter, likely larger still if we follow the emission down to the noise of the map ($\sim$\,0.9$\,\times\,$10$^{19}$\,atoms\,cm$^{-2}$).  At the center of this gas cloud, there are no detections in any other band
(Figure ~\ref{fig:4plot}b, c, d), even at the faintest flux levels, or using aggressive smoothing to increase the S/N of any faint emission. This is readily apparent if we zoom into the cloud core using the deep KiDS imaging, see Figure ~\ref{fig:1plot}.

The deepest optical imaging we have available are from the KIDS DR4 \citet{Kui19}, with \textit{r}-band magnitudes reaching 25.2\,mag (AB; 5\,$\sigma$) in a small aperture due to the high-quality sub-arcsec seeing.  The `\textit{g}' and `\textit{i}' bands have similar limiting magnitude depths.
In the zoomed color composite (Fig.~\ref{fig:1plot}), we see a number of small sources, similar in number density to the field outside of the cloud region. None have matches in the GAMA spectroscopic catalogue (see below), so their true nature is not certain.  We can say that none are in close proximity to the \HI\ core ($>$20$\,\times\,$10$^{19}$\,atoms\,cm$^{-2}$) of Source (b).  
We are able to   recognize disk and ellipsoid galaxies, both their size and red (passive) colour indicating likely background galaxies.  On the periphery of Source (b), near the outer \HI\ contours, there are faint and compact blueish sources that may be star-forming galaxies that are distant, or dwarf galaxies in proximity to the $z=0.03$ filament, or most intriguingly, may be SF complexes or knots associated with the \HI\ gas cloud.  The most promising blue blob to follow up is at 22h56m20s -33d58m08s (J2000), ideally with spectroscopy, but in more practical terms, 
deeper optical imaging should be carried out to reveal if there other knots or blobs that would indicated an ultra-diffuse galaxy in proximity to Source (b).

We have made use of a photometric redshift catalogue to investigate the sources in the vicinity of the cloud chain.
We find no obvious associations with the cloud or filament at $z = 0.03$.  GAMA provides EAZY \citep{Brammer08} photometric redshift estimates for all optical/NIR detections \citep{Bellstedt20}, which are based on fits to the $u$--$K$ SEDs (for details see Driver et al 2021, under review). 
For the sources near the central cloud core (Fig.~\ref{fig:1plot}, within the 3 inner contours, or $>$5$\,\times\,$10$^{19}$\,atoms\,cm$^{-2}$),
the GAMA DR4 photo-z's range from 0.3 to 0.6, which very likely rules them out as potential associations even given the 10\% uncertainty in the photo-z's.

Considering the filament galaxies, the nearest Group-83 galaxy members
are indicated in Figure~\ref{fig:meerhogs_massive_dark_cloud_enviro}, located within 500\,km\,s$^{-1}$ of the Cloud central velocity;  most notably, the bright  \textit{I}, \textit{II} and \textit{III} galaxies (see also Fig. ~\ref{fig:4plot}b and Figure~\ref{fig:1plot}, and properties in Table~\ref{tab:enviro_prop}).
Nearest, galaxy \textit{I} is about 2\arcmin\ from the cloud core, and its velocity is such that it would be located at least 70\,kpc distant from the cloud core (barring peculiar motions).  Their association with the cloud may only be gravitational; they do not appear to be ``hosts" for the cloud at the current epoch that is observed.  Except for \textit{III} which is a small disk galaxy, they are gas poor and passive in star formation properties (i.e., they are old, evolved galaxies).  However, given their proximity to the cloud chain, past tidal interaction may have occurred, and moreover \HI\ gas may have been disruptive to later form the clouds.  We discuss such a scenario below (Section \ref{sec:discussion}), notably comparing with the Leo Ring system.\\

We tentatively conclude that the dark cloud is not detected at wavelengths that we associate with extragalactic stellar and ISM emission. The imaging does, however, provide an estimate for an upper limit flux for the detection, from which we may then examine the implied ratio between the gas mass and the stellar luminosity 
(i.e., $M_{\HI}$ mass versus optical or infrared luminosity).  

Here we consider the deep \textit{r}-band image (Fig.~\ref{fig:4plot}, upper right panel, and Fig.~\ref{fig:1plot}).   In the vicinity of the dark cloud, we measure a background 1-$\sigma$ noise of 25.6\,mag\,arcsec$^{-2}$ (0.21\,$\mathrm{\mu}$Jy\,mag\,arcsec$^{-2}$), 
which translates to a 
3-$\sigma$ detection of 24.4\,mag\,arcsec$^{-2}$ considered over a 25\,arcsec$^{-2}$ area.
At the luminosity distance of the dark cloud, 126\,Mpc, the corresponding absolute magnitude is -11.1\,mag, corresponding to a luminosity of 10$^{6.3}$\,L$_\odot$ (where we have used a zero point AB magnitude of 4.65).  This would be a very low mass, dwarf galaxy.  
Now comparing this 3-$\sigma$ luminosity limit to the \HI\ mass of the dark cloud (b), 10$^{9.70}$\,$M_\odot$, the implied M$_{\HI}/L_{r}$ is at least $\sim$2600.  
Even at a higher detection limit, say 5-$\sigma$ surface brightness (23.8\,mag\,arcsec$^{-2}$), the implied
M$_{\HI}$/L$_{r}$ is greater than 1000.  

Consider the mass ratio for Source (b).  Using the 3-$\sigma$ \textit{r}-band luminosity and a dwarf galaxy stellar M/L = 2 \citep{Mar08, Flynn06}, the limiting stellar mass is 10$^{6.6}$\,L$_\odot$, and the ratio between the neutral gas and the stellar content, M$_{\HI}$/M$_{\star}$, is 10$^{3.1}$. The extremely high mass ratio is well outside of any scaling relation, and is an order of magnitude larger than the estimated value for the recently discovered massive dark cloud, AGC\,229101 \citep{Leisman21}.  We note that the stellar M/L for dwarf disks and spheroidals may range widely, over an order of magnitude;  conservatively using a range between 1-10 \cite[e.g., see ][]{Flynn06}, the corresponding range in M$_{\HI}$/M$_{\star}$ is 10$^{2.4}$-10$^{3.4}$.

Given the absence of a detection, and if we assume the dark cloud is a galaxy with a past history, we can then expect it to be very low mass and low surface brightness, and likely well resolved by the KiDS imaging.  An ultra-diffuse galaxy (UDG) would have an angular size that ranges from a $\sim$few to 13\arcsec\ at the distance to the dark chain (see e.g., the UDG size distribution in \citealt{van_dokkum_forty-seven_2015} or the sizes of \HI\--bearing UDGs in \citealt{leisman_almost_2017}).  Even comparing to the classic LSB galaxy, Malin-1, it has a central surface brightness close to 26\,mag\,arcsec$^{-2}$ \citep[in the V-band;][]{Bothun87}, slightly fainter than our KiDS image sensitivity (note however, the correlated pixels from such a large object would bin up and be easily detected with KIDS).

Nevertheless, binning up the images to effectively smooth the background and peak up any diffuse emission lurking in the noise, reveals nothing at the location of the dark cloud, not even a hint of a faint smudge.  Similarly in the \textit{g} and \textit{i} bands.
It is fair to say this large gas cloud is extremely dark in comparison to the optical emission.  There may in fact not be any past star formation.  A deeper optical image, reaching at least 27 to 28\,mag\,arcsec$^{-2}$ is in order to probe for even lower mass systems.

At the longer wavelengths, the situation does not change.  Consider the mid-infrared 
WISE W1 image; Figure ~\ref{fig:4plot}b.
It has a 3$\sigma$ surface brightness of 22.7\,mag\,arcsec$^{-2}$ (Vega; 0.26\,$\mathrm{\mu}$Jy\,arcsec$^{-2}$), which is also very faint and implies a high mass-to-light ratio.   Estimating the ratio using a more conservative 5$\sigma$ limit of 22.2\,mag\,arcsec$^{-2}$) (Vega; 0.4\,$\mathrm{\mu}$Jy\,arcsec$^{-2}$), the absolute magnitude for this limit is -13.3 mag (Vega), 
and the corresponding in-band luminosity (adopting a zero point magnitude of 3.24, see \citet{Jar13}) is 10$^{6.6}$\,L$_\odot$ (which also implies a low stellar mass, 
$\sim$10$^{6.3}$\,M$_\odot$).
The implied $M_{\HI}$/L$_{\textrm{IR}}$ is $\sim$1300, similarly extreme as what was found in the optical comparison. 

%

\subsection{Associations for the H {\sc i} Chain Sources?}

It is clear that the central concentration to the chain of \HI\ detections is not obviously detected in the UV, optical, or infrared, making it a massive, truly dark cloud object (Fig~\ref{fig:1plot}).  Here we examine the other sources detected in \HI\,  denoted as Sources (a)-(g); see  Table~\ref{tab:HIprop} and Figure~\ref{fig:1plot}.   

Source (a) has a peak column density of 8.8$\,\times\,10^{19}\,\mathrm{atoms}\,\textrm{cm}^{-2}$ and a total atomic hydrogen mass of 10$^{8.88}$\,$M_\odot$.  It has a lower peak column density compared to the main cloud, and appears much more diffuse in nature.  At this central position, and within arcminutes radius, there are no optical, infrared or UV detections whatsoever (the nearest GAMA galaxy is labeled \textit{V} in Figure~\ref{fig:meerhogs_massive_dark_cloud_enviro}, a non-group member at the more distant redshift of 0.0324), which also implies this is a very dark patch of the cloud chain.

Source (b) is the massive concentration (see above for discussion of the properties and M/L estimates).\\

\begin{figure}
\includegraphics[width=0.49\textwidth]{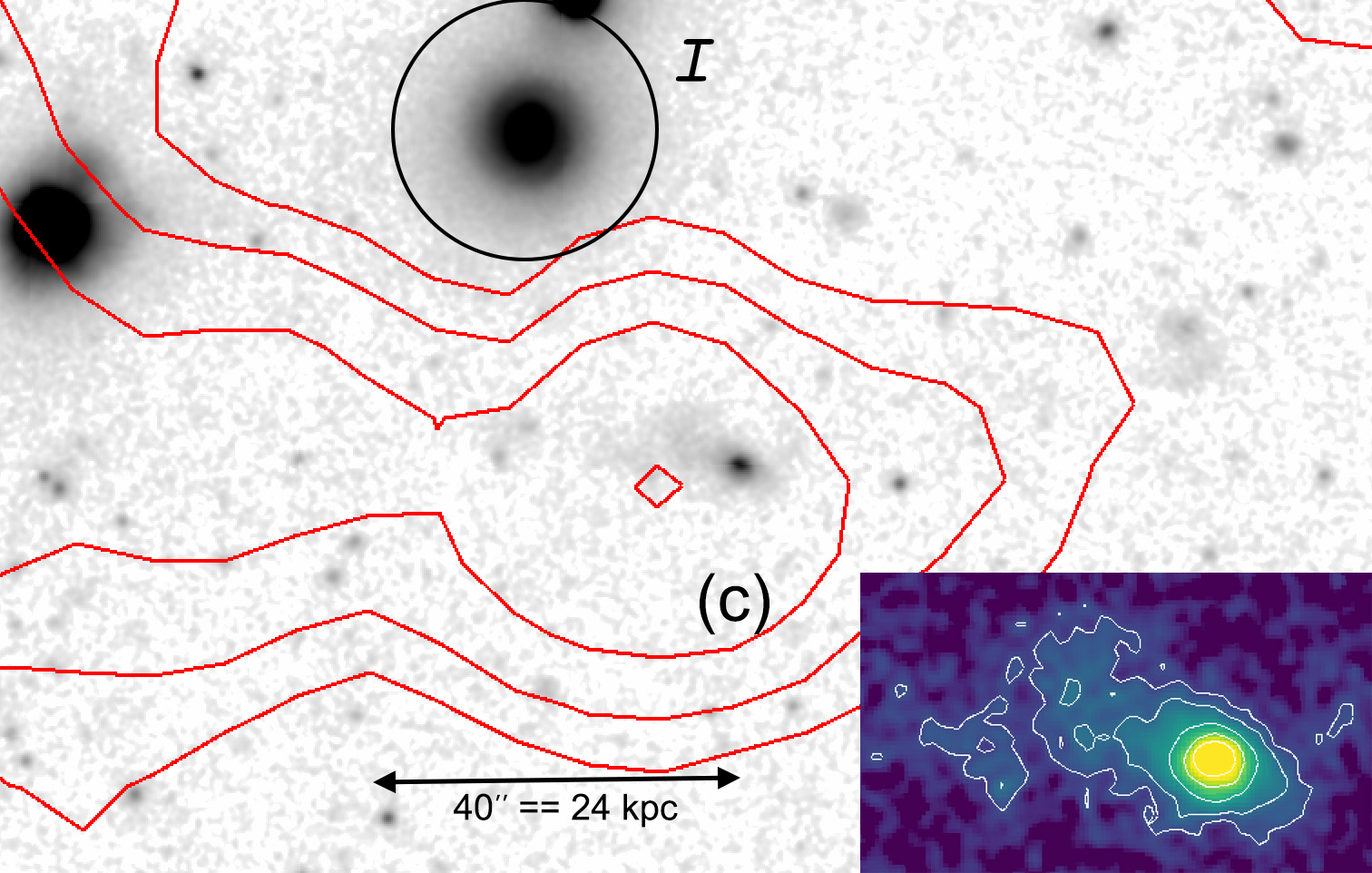}
\caption{Close view of Source (c) with KiDS \textit{r}-band imaging.
As in the previous figures, the \HI\  contours (red) indicate
the gas distribution, with the (c) concentration in close proximity to the optical galaxy.
Shown in the lower right is an inset image of a zoomed view of the optical source (with colour transform ``viridis"), located at (deg J2000) 344.03668 -33.961783. The white contours, ranging from 25.8 to 22.5\,mag\,arcsec$^{-2}$, delineate the low surface brightness `tidal tail' extending to the eastward toward the massive dark cloud core.  
}
\label{fig:westsrc}
\end{figure}

Potential Association:  Source (c) is located directly west of the central mass (Source b) and 55\arcsec\ due south of Group-83 member \textit{I} (Fig.~\ref{fig:4plot}) and Figure~\ref{fig:1plot}. 
It appears to be connected through a bridge of gas that extends $\sim$2\,arcmin ($\sim$\,70\,kpc) from the cloud core (Fig.~\ref{fig:4plot}a), implying a physical association with the main cloud core.  Intriguingly, there is a faint optical source about 13\arcsec\ to the west of the central location of the \HI\ detection we label as (c); Figure~\ref{fig:4plot}b.  This is a plausible association, well within the HWHP beam of this \HI\ map.

Moreover for Source (c), close inspection of the \textit{r}-band image (see Fig.~\ref{fig:westsrc}) reveals that the optical source has a distorted `tail' that extends north-eastward towards the massive \HI\ cloud (Source b).  Might this be a gravitational tidal distortion coming from current or previous group interaction?  Integrating a large (20\arcsec\ diameter) aperture to capture the core and tail emission, the resulting AB magnitude is 19.51$\pm$0.01, corresponding to a flux density of 57.0\,$\mathrm{\mu}$Jy.  There is no optical GAMA redshift for this source, so we cannot be certain it is associated. The reason this source does not have a GAMA redshift is because in the original GAMA target catalogues, which are based on AUTO photometry of SDSS imaging, it was fainter than the 19.8 mag \textit{r}-band selection limit.  It does, however, have a GAMA DR4 photometric redshift because it was detected and extracted in the deeper photometric catalogues.  The GAMA source is located at RA, DEC (J2000\,deg) = 344.037096, -33.96165.
For this source, the best photo-$z$ estimate is $z_{peak} = 0.0357$ with a 68\% confidence interval $(0.024, 0.049)$, which is certainly consistent with this galaxy being a member of Group 83, or at least the z=0.03 filament.

If we thus consider this to be an association, and adopt the distance of Source (c), the  optical \textit{r}-band luminosity would be 10$^{8.30}$\,L$_\odot$.  The corresponding \HI\ mass for this source is 10$^{9.15}$\,$M_\odot$ (see Table~\ref{tab:HIprop}), which would then imply a M$_{\HI}$/L$_r$ of $\sim$8.  This is a plausible ratio (although on the high side) for 
dwarf and star-forming galaxies \citep{LSS92, Nal21}, while also more extreme metal-poor dwarf examples have similarly high ratios \citep[see e.g.,][]{Filho13}.
In addition, Source (c) is also clearly detected in the UV (Fig.~\ref{fig:4plot}d), which strongly suggests active star formation. This is consistent with a tidal or a gravitational encounter,  triggering recent star formation.

Other bands:   At lower energies,  the source is detected in the WISE near-IR (W1) image (Fig.~\ref{fig:4plot}c), at coordinates[deg] 344.03632 -33.96150.  Emission at this wavelength implies an older stellar population is already in place.  We can estimate the stellar mass using the prescription from \citet{Clu14}; starting with the integrated W1 magnitude, 17.16$\pm$0.11 ($\sim$42\,$\mathrm{\mu}$Jy), and assuming the distance to the cloud is 126 Mpc, the in-band luminosity is then 10$^{8.64}$\,L$_\odot$. Adopting a M$_{\star}$\,/\,L ratio of 0.4, appropriate to late-type galaxies, the implied stellar mass is then 10$^{8.2}$\,$M_\odot$.  

Hence, if this source is associated with the \HI\ cloud,
 namely Source (c), it is a dwarf galaxy that has recently been activated in star formation, likely from tidal forces, and which possesses older stellar populations anchoring the internal structure.  The gas-to-stellar mass ratio (M$_{\HI}$/M$_{*}$) is $\sim3$, which is well above the typical scaling relations seen for nearby galaxies, with ratios $\le$1 \citep[see eg.][]{Nal21}.  We surmise either too much gas is attributed to this clump (since it is part of a larger complex, with a gas bridge to its massive neighbor), or the cloud is anomalously rich in neutral hydrogen.
 \\

Moving to the north side of the cloud complex, 
Source (d) is to the NW of the massive cloud (b), and is directly 1\arcmin\ north of Group-83 member \textit{I} (note that both Source c and d are about equidistant N-S from Group-83 member \textit{I}).   This source also  
appears to be connected by a gas bridge to the larger complex; see Fig.~\ref{fig:4plot}a/b.  In close proximity to the \HI\ detection, we see a clear optical detection, roughly 8\arcsec\ from the nominal \HI\ centroid (i.e., within the beam of the radio measurements).  There is no optical redshift for this source, so we cannot definitively associate it with the gas concentration.  If we make the assumption that it is the host of the Source (d) gas emission, we can estimate its stellar mass using the WISE W1 image, also well-detected (Fig.~\ref{fig:4plot}c).  The source is seen in WISE at J2000 coordinate[deg] 344.04672 -33.93435, and has an integrated W1 magnitude (Vega) of 15.31$\pm$0.31 (0.23\,mJy), and W1-W2\, color of -0.03\,mag, and is undetected in the longward (ISM-sensitive) W3 and W4 bands.   These color properties imply an early-type, likely bulge-dominated, host galaxy.  The estimated W1 luminosity and stellar mass follows from assuming the distance (126\,Mpc) to the cloud and the M/L color method of \citet{Clu14}:  $10^{9.36\pm0.03}$ L$_\odot$ and 10$^{9.1}$\,$M_\odot$, respectively. Hence, if this source is associated with the cloud complex, then the computed stellar mass implies a dwarf spheroidal galaxy.  This possible association seems less certain given the early-type colors of the optical/IR sources, which is more consistent with a gas-poor galaxy; nevertheless, it is a plausible association.  

Moving to the last set of sources in the cloud complex:  to the west of the massive gas core, Source (e) has no obvious large counterpart in the UV-optical-infrared.  But there is an faint blue fuzzy source $\sim$10\arcsec\ to the east, located at (J2000): 344.01206\degr -33.94493\degr, which is worthy of follow-up.  Similarly but on the opposite side of Source (e) to the west, there is diffuse emission that is well spread across 20 to 30\arcsec in length.  As best we can measure, the center of this diffuse structure is located at 344.0040\degr, -33.9458\degr, and overall has a blue color implying active or recent star formation.
It is plausible that this is a very low SB galaxy lurking near or within the \HI\ cloud chain, or is a piece of a larger UDF galaxy system, 
and is worthy of follow-up deep imaging.

Source (f), located in the westernmost cloud which is closest to the Group-83 center (compact group of galaxies), seems to split a close pair of faint optical galaxies (Fig.~\ref{fig:4plot}b). Fortunately, these optical counterparts do have a redshift, but with a value of $z\sim0.16$, i.e. they are well distant galaxies and hence clearly not associated with the \HI\ complex.  

Source (g), the final source of the gas chain is in the 
westernmost gas concentration.  It has a more amorphous and diffuse morphology, less certain to be a discrete galaxy.  There are no UV-optical-infrared sources in close proximity to its core.  However, there is an early-type galaxy about 30\arcsec\ to the south-west of the core (see Fig.~\ref{fig:4plot}b, green circle near Source g) . This source is identified as being in the filament and part of a galaxy pair (N=2 group).
It has a redshift of $z$\,=\,0.0272 (vs. 0.0294 for (g)), and hence is slightly in front of the cloud complex (115 vs 125 Mpc, assuming no peculiar motions).  This relatively bright source (W1\,=\,13.38\,mag or 1.4\,mJy), located at (deg J2000) 343.9182; -33.9426, has WISE colors that indicate it is an early-type spheroid, with little or no star formation activity.   Given the distance, velocity and color mismatch,  it is therefore unlikely to be associated with the \HI\ dark cloud chain.  

Finally, we note that Sources (d)-(g) appear to have a more filamentary structure, less discrete morphology, and are aligned toward the Group's compact center, located 9 to 10\arcmin ($\sim$
$\sim$310 -- 350 kpc) further to the west, with clouds (f) and (g) being connected by a gas bridge. This connection (f)-(g) shows the clearest velocity gradient of the whole complex (without having any clear association with a galaxy).

%

\section{Discussion}\label{sec:discussion}

Three properties of the dark cloud chain are notable:  (1) it is a dark \HI\ complex, with a central source that is as massive as a gas-rich galaxy, but which shows no signs of previous or current star formation (yet one of the smaller concentrations may be associated with a tidal-distorted dwarf galaxy). Further to the absence of star formation activity, we note that radio continuum emission, at the position of the dark cloud chain, is not detected.  On the other hand, the \HI\ emission is detected in the single-dish HIPASS survey, with a peak surface brightness and integrated flux that is consistent with the MeerKAT measurements (see Appendix~\ref{appendix_c}), thus adding confidence of a real astrophysical detection. 
(2) It has hardly any kinematic structure along the radial direction, no larger than about 110\,km\,s$^{-1}$, and with little evidence of rotation, and (3) it is located on the extreme edge of a large galaxy group:  Group-83,  $N=18$ members with $z_\mathrm{fof}=0.0289$ and a dynamical halo mass of 10$^{13.46}$\,$M_\odot$ and velocity dispersion of 228.6 km\,s$^{-1}$.

The cloud complex also appears `stretched' with multiple concentrations 
 in the direction of the dense compact core of the Group-83.  There is a noticeable absence of \HI\ gas in the Group, other than the cloud complex itself;  it is a gas-poor group otherwise.  What is the origin and nature of this massive cloud chain.  Is the gas accreting into the Group, perhaps merging from smaller clouds and now just entering the outer boundary, or might it be the ripped-apart detritus from an earlier tidal event between gas-rich massive galaxies in the group? That would require a very massive cast-off, from hosts that are not readily apparent in the region or even the filament itself.  If this cloud is not self-supporting, how can it persist against tidal and photo-disruption?  We do not have enough information at this time to fully address these compelling questions.

As reviewed in Section \ref{sec:introduction}, these types of dark \HI\ clouds are quite rare, and massive ones ($>10^9$\,$M_\odot$) are rarer still. Here we have a complex that has a discrete source as massive as 10$^{9.7}$\,$M_\odot$, and if we combine the gas for all seven sources the aggregate mass is $10^{10}$\,$M_\odot$, a rather large amount of neutral hydrogen floating at the edge of a sizable galaxy group (with a compact core)  
that is 3.5 orders of magnitude more massive (in total) than the gas cloud complex itself.  Where did this neutral hydrogen come from, where is it going and why is it not forming stars?

Notable examples of massive \HI\ associations are the Leo\,Ring \citep{schneider_discovery_1983,schneider_neutral_1985,schneider_high-resolution_1986,schneider_neutral_1989,stierwalt_arecibo_2009}, with a recently studied analogue AGC203001 \citep{bait_discovery_2020}, HI\,1225+01 \citep{giovanelli_protogalaxy_1989}, Hanny's Voorwerp \citep{lintott_galaxy_2009,jozsa_revealing_2009,keel_history_2012}, HCG~44 \citep{serra_discovery_2013,2017HessCluver} and AGC\,229101 \citep{Leisman21}. 

The Leo\,Ring is an \HI\ ring with a diameter of $200\,\mathrm{kpc}$ and a total mass of $10^{9.3}\,M_\odot$ in the M96 galaxy group, long thought to lack a stellar counterpart (\citealt{schneider_high-resolution_1986}). As with the dark cloud chain, the ring forms a coherent, diffuse envelope around several concentrations. More recently it has been shown that the Leo\,Ring contains faint emission in the UV \citep{thilker_massive_2009}, and/or has dust emission \citep{bot_search_2009}, and/or emission in the optical \citep{michel-dansac_collisional_2010} in a low number of concentrated \HI\ clumps. Additionally, the metallicity measured along two sightlines is estimated to be at $1/10\,Z_\odot$, showing that the Leo Ring, if of primordial origin, has to be enriched by the ISM of surrounding galaxies and - if tidal - must stem from a low-metallicity object, making the disruption of an LSB galaxy a potential scenario for its formation \citep{rosenberg_unraveling_2014}. This would, however, also result in the additional detection of a stellar component. Stars have been associated to knots in the ring \citep{michel-dansac_collisional_2010}, but a search for more massive dwarf objects has borne ambiguous results so far (\citealt{stierwalt_arecibo_2009}, see also \citealt{bait_discovery_2020} for a discussion). 

As an association of \HI\ concentrations, the Leo Ring can be compared to the whole dark cloud chain or associations therein, rather than just one of its constituents. The dark cloud chain differs from the Leo Ring in a number of important ways.  First, we identify a star forming dwarf galaxy as potentially being associated to Source (c). Second, the total mass of the dark cloud chain with ($10^{10.0}\,M_\odot$) is 5 times larger than the Leo Ring and, third,  it is twice as extended (as the diameter of the Leo Ring equals half the linear size of the dark cloud chain). At first sight, one might interpret the appearance of the dark chain as a chain with two linear parts, (a)-(c), and (d)-(g) rather than a ring. On the other hand, considering the configuration of Sources (b)-(e) and the surrounding \HI\ only, with a diameter of $3.\!\!\arcsec6\,\equiv\,125\,\mathrm{kpc}$ between Sources (b) and (e), the impression arises that the inner part of the observed \HI\ might form a ring with a projected major axis between Source (b) and Source (e). Sources (a), (f), and (g) would then constitute extensions, which are also observed for the Leo Ring \citep[towards M96, see][]{schneider_neutral_1985}. With the early-type Group-83 galaxy \textit{I}, the dark cloud chain (or ring) might also have a central object.

However, unlike the Leo Ring,  the dark cloud chain would barely rotate, as the velocity difference between (b) and (e) with $\sim 20\,\mathrm{km}\,\mathrm{s}^{-1}$ lies below the nominal velocity resolution of our data. Correcting for an inclination of minimally $\sim 30^{\circ}$ (by inspection) results in a rotational amplitude of $\lesssim 20\,\mathrm{km}\,\mathrm{s}^{-1}$. The Leo Ring has a rotational amplitude of $\sim 250\,\mathrm{km}\,\mathrm{s}^{-1}$ \citep{schneider_neutral_1985}, which differentiates it from the cloud chain. Moreover, the systemic velocity of Group-83 galaxy \textit{I} differs from that of the dark cloud chain by $\sim\,500\,\mathrm{km}\,\mathrm{s}^{-1}$, making an association somewhat tenuous. Given the limited kinematic information of our MeerKAT observations, we cannot rule out a ring-type gas distribution, perhaps associated with a galaxy-galaxy collision (see below),  yet the properties of the cloud and the local environment do seem to indicate a different framework in place.

Consider a similar ring-like object, a massive ring around the galaxy AGC~230001 with a diameter of $115\,\mathrm{kpc}$, an \HI\ mass of $M_{\HI}\,=\,10^\mathbf{9.4}\,M_\odot$ and with little stellar mass has been found by \citet{bait_discovery_2020}. Interpreting the velocity gradient along the AGC~230001 ring ($W_{20}\,=\,100\,\mathrm{km}\,\mathrm{s}^{-1}$) as rotation, and assuming an inclination of $i\,\sim\,60^\circ$ the estimated rotational speed of $\sim\,60\,\mathrm{km}\,\mathrm{s}^{-1}$ is still significant, clearly larger than the case for the dark cloud chain. As with the Leo Ring, it too is kinematically and morphologically associated to a (quenched) massive galaxy (AGC~230001), which has the same systemic velocity as the ring). 

It might hence appear that the Leo Ring or AGC~230001 differs from the dark cloud chain. Nevertheless, at least attempts to explain their nature might be applicable to the dark cloud chain. \citet{michel-dansac_collisional_2010} simulate a head-on collision between two gas-rich spiral galaxies, resulting in a ring-like structure of extended debris, that the authors compare to the Leo Ring. Candidates for galaxies in this type of scenario in the case of the dark cloud chain could be Group-83 galaxies \textit{I} and \textit{II}, both of which are gas-poor and passive in SF. The Leo Ring and AGC~230001 will remain objects for comparison in the future as the data on our newly discovered object accumulates.

Continuing our review of remarkable dark \HI\ systems, 
H\,1225+01 is an enigmatic system at the edge of the Virgo Cluster, consisting of a giant \HI\ cloud with two massive concentrations,
 a NE concentration with $M_{\HI}\,=\,10^{9.4}\,M_\odot$, and an SW concentration with a mass of $M_{\HI}\,=\,10^9\,M_\odot$ respectively), the more massive of which harbours a diffuse star-forming dwarf galaxy, and for the less massive of which no stellar counterpart has been detected \citep{djorgovski_preliminary_1990, mcmahon_optical_1990, giovanelli_complete_1991}. \citet{matsuoka_updated_2012} showed that the NE cloud has ongoing star formation, while the SW concentration does not
contain stars to a limit of $\mathrm{R_{AB}}\,>\,28\,\mathrm{mag\, arcsec}^{-2}$ 
over an area of 10\arcsec.

The \HI\ radius of HI\,1225+01 compared to its optical extent would be outstanding among dwarf galaxies \citep{hoffman_arecibo_1996}. The two clouds are embedded in a common envelope of \HI. This system might be the most similar to the dark cloud chain, although it has half the mass and half the extent ($\sim 200\,\mathrm{kpc}$) of the dark cloud chain. It shows a shallow velocity gradient of ~$44\,\mathrm{km}\,\mathrm{s}^{-1}$ between the two concentrations, while the spectral width of the concentrations itself is $\sim 45\,\mathrm{km}\,\mathrm{s}^{-1}$ for the NE cloud and $\sim 25\,\mathrm{km}\,\mathrm{s}^{-1}$ for the SW cloud \citep{giovanelli_complete_1991}. Similar velocity gradients in the dark cloud chain would be consistent with our observations with their coarse velocity resolution. Again, whether the dark cloud chain and its largest concentration, Source (b), might be the more massive sister system to HI\,1225+01 can only be confirmed in future studies, in particular high-resolution \HI\ observations.

Hanny's Voorwerp is a large ionization nebula in the surroundings of the disk galaxy IC~2497, separated by $\sim 20\,\mathrm{kpc}$ \citep{lintott_galaxy_2009}. \citet{jozsa_revealing_2009} could show that the ionization region is part of a massive \HI\ cloud with $M_{\HI}\,=\,10^\mathbf{9.7}\,M_\odot$, which itself has an additional companion some $100\,\mathrm{kpc}$ to the West of IC~2497 with a mass of $M_{\HI}\,=\,10^\mathbf{9.5}\,M_\odot$, again potentially connected by a gas bridge. The galaxy itself was not found to contain significant amounts of \HI. While the literature largely concentrates on the origin of the ionisation nebula as a light echo of a ceased quasar at the centre of IC~2497, the origin of this extremely massive extragalactic cloud system has not been studied in detail. \citet{keel_history_2012} showed that the ionization nebula shows star formation, probably induced by a mild shock from an outflow of IC~2497, but no additional stellar component to the rest of the cloud. While the system is hence comparable to the dark cloud chain in mass and in extent, it is insofar different as an interaction partner, IC~2497 is clearly present. Should the cloud complex be the result of a tidal interaction, though, the whereabouts of the interaction partner would not be known, given the lack of follow-up studies.

Hickson Compact Group HCG~44 might provide a clue to the potential origin of massive, seemingly isolated clouds. As first discovered by \citet{serra_discovery_2013}, HGC~44 shows a very elongated \HI\ tail, which could be the result of intra-group gas stripping or a tidal interaction of the group with the spiral galaxy NGC 3162 at a distance of ~$650\,\mathrm{kpc}$. No stars were found to a limiting \textit{g}-band surface brightness of $28.5\,\mathrm{mag\, arcsec}^{-2}$. \citet{2017HessCluver} demonstrated that the extent of the tail is $450\,\mathrm{kpc}$ with a total mass of $10^9\,M_\odot$. While this is a factor of 10 less than the dark cloud chain, it is apparently possible to generate extremely massive, extremely wide-spread debris in galaxy interactions.
In the case of the dark cloud chain, should it be the result of tidal origin, we might venture an age estimate by assuming an encounter at the centre of the central group, at a distance of $400\,\mathrm{kpc}$, assuming an upper limit of twice the dispersion of the group ($457\,\mathrm{km}\,^{-1}$) as its velocity, resulting in a lower limit of $700\,\mathrm{Myr}$.

Might there be an extreme low surface brightness galaxy, akin to UDGs, just beyond the sensitivity of our optical and infrared images?  UDGs, with a central surface brightness $>$24 mag\,arsec$^{-2}$ (AB) (and r$_{eff} > $1.5\,kpc) are found in galaxy groups and cluster environments at the distance of Group-83 and
the dark cloud chain \citep[see e.g.,][]{van_dokkum_forty-seven_2015,VDB17,ZAR19}; moreover, KiDS imaging has been successfully deployed to find
such objects down to 26\,mag\,arsec$^{-2}$ \citep{Karademir21}. For a group with  mass the size of Group-83, we can expect to find a few UDGs ($\sim$4 to 5) based on number counts versus group M$_{200}$ masses \citep[see Fig.3 in][]{VDB17}.  Our \textit{r}-band image is sensitive enough to detect UDGs down to $<$26\,mag\,arsec$^{-2}$ or so, which is more than adequate to detect prototypical UDGs.  And yet inspection of the central region of the dark cloud chain does not reveal any such faint emission, nor with additional binning and smoothing to enhance signal-to-noise.  Either the stellar content is even lower mass and fainter than these limits, or the dark cloud is not forming, nor ever has, formed stars --- this seems unlikely if the Source (c) association with a dwarf galaxy is real, and the diffuse blue-fuzzy source close to Source (e) is also associated.

A recent dark cloud discovery \citep{Leisman21} of a similarly massive neutral hydrogen cloud has identified a dwarf mass stellar counterpart at a surface brightness that is 0.5 to 1.0 mag below our own \textit{r}-band limit, which emphasises the possibility of an  older stellar population at such extreme surface brightnesses, which would be in line with our $M_{\HI}/M_{\star}$ estimate being greater than 1000.

A final consideration, the dark matter content of the cloud.
If the gas cloud is potentially relaxed, would it require dark matter to stabilize? Using the line width w$_{20}$\,=\, 139\,$\mathrm{km}$\,$^{-1}$ and assuming a Gaussian line profile, we derive a dispersion $\sigma_\mathrm{z}\,=\,$39$\,\mathrm{km}\,\mathrm{s}^{-1}$.
Assuming a diameter of $2R\,=\,50\,\mathrm{kpc}$ and a constant density distribution, the virial mass $\frac{M_\mathrm{Vir}}{M_\odot}=2.325\,\times\,10^5 \,\times\, 5\left(\frac{\sigma_\mathrm{z}}{\mathrm{km}\,\mathrm{s}^{-1}}\right)^2 \times \frac{R}{\mathrm{kpc}}$ of the central concentration is 4.4$\,\times\, 10^{10}\,M_\odot\,=\,10^{10.6}\,M_\odot$.
With the assumption of an isothermal ($\frac{1}{r^2}$) mass profile ($\frac{M_\mathrm{Vir}}{M_\odot}=2.325\,\times\,10^5 \,\times\, 2\left(\frac{\sigma_\mathrm{z}}{\mathrm{km}\,\mathrm{s}^{-1}}\right)^2 \times \frac{R}{\mathrm{kpc}}$), this reduces to 1.8$\,\times\, 10^{10}\,M_\odot\,=\,10^{10.2}\,M_\odot$ \citep[see e.g.][]{hoffman_arecibo_1996}. 
Assuming a full rotational support, a lower limit (assuming an inclination of $90^\circ$) to the dynamical mass ($\frac{M_\mathrm{dyn}}{M_\odot}=2.325\,\times\,10^5 \,\times\, \left(\frac{0.5\, w_{20}}{\mathrm{km}\,\mathrm{s}^{-1}}\right)^2 \times \frac{R}{\mathrm{kpc}}$) is 2.8$\,\times\, 10^{10}\,M_\odot\,=\,10^{10.4}\,M_\odot$.
This means that Source (b) with an \HI\ mass of $M_{\HI}\,=\,10^{9.7}\,M_\odot$, should it be a stable structure (whether rotationally or pressure-supported), likely requires some amount of dark matter to be long-lived, with a total-to-luminous matter ratio of 2.6$\,{>}\frac{M_\mathrm{tot}}{M_\mathrm{lum}}$
(scaling the \HI\ mass by a factor of 1.4 to account for helium). Without any rotation this is less than would be typical for a galaxy. Notice that no attempt has been made to correct the line widths for the effect of instrumental broadening, as the filter function of the MeerKAT bandpass is close to a rectangle (nearly no spill from one channel into the neighbouring channels). A slight broadening would bias our results to a higher total mass, underlining the conclusion that Source (b) might have a deficiency of dark matter as compared to a galaxy. At this stage it is, however, not known whether the cloud complex has any gravitationally bound concentrations, such that it is well possible that their kinematics and mass distribution are not obviously connected.  Moreover, given the implied dynamical or halo mass we would expect a dwarf galaxy of  M$_{*}\ge10^9$\,M$_\odot$, which is not the case.

The cloud chain remains an enigma, with many open questions;
it will take further investigation to address these issues, starting with higher velocity resolution observations to better understand the kinematics of the complex, and deeper optical imaging to detect or rule out an extreme LSB galaxy. Notwithstanding the total absence of past star formation, a reasonable hypothesis to test is the possibility of a large disk galaxy being viewed relatively face-on, with smaller cloudlets (accretion?) associated with the disk and with similar kinematics in the plane of the sky.

To shed more light into the nature of the cloud complex and the enigmatic brightest concentration, we will pursue  full 32K observations using MeerKAT.  Moreover, 
Cluver rt al. (in preparation) will fully address the properties of the galaxies inhabiting the greater filament that houses the chain, including the star formation and stellar mass properties of the larger group, to help understand the environment in which the dark cloud chain has been evolving.

%

\section{Summary}\label{sec:summary}

The MeerKAT Habitat of Galaxies Survey (MeerHOGS) was created to map local large scale structures to investigate the gas-to-star formation processes, central to galaxy evolution studies.  Another purpose of MeerHOGS was to test preliminary (early development) of \HI\ pipelines for the MeerKAT radio telescope.  A prominent filament in the GAMA G23 Region was targeted, covering $\sim$7\,deg$^2$ and a redshift range from 0.026 to 0.034 (7800 to 10200\,km\,s$^{-1}$ in c$z$).  

We detected over 60 sources in \HI, spread throughout the filament ranging in \HI\ gas mass from $10^{8.27}$\,to\,$10^{9.92}$\,$M_\odot$.   But in the course of comparing the \HI\ sources to our GAMA and 2dFGRS redshift catalogues, we identified a `chain' of \HI\ emission with a total \HI\ mass of $10{^{10}}$\,$M_\odot$ that did not have any counterparts.  Further investigation, using deep ultraviolet, optical and infrared imaging (GALEX, KiDS and WISE, respectively), reveal this gas chain to be extremely dark: there is no emission from stars or star formation in proximity to the dark chain, most notably the massive central source, with $10{^{9.7}}$\,$M_\odot$ in neutral hydrogen within 50\,kpc diameter, implying an extreme gas mass-to-light ratio.  One smaller source located $\sim$70 kpc to the west of the cloud core, however, may have a faint optical counterpart which has dwarf (low-mass) properties and is tidally disturbed (assuming it is associated and hence at the distance of the cloud complex); this is likely given the photometric redshift consistent with being at the distance of the cloud or large-scale filament.     

In general, the cloud complex appears to be  
seven \HI\ peak concentrations, extending $\sim$400 kpc
from east-to-west (and possibly larger, smaller clouds are seen further afield yet aligned).   The source has been validated to be real and not an artifact, and moreover, it is detected in HIPASS with a peak surface brightness and integrated flux that are wholly consistent with the MeerKAT measurements.   

The size and mass of the central concentration is expected for a large, gas-rich, rotating galaxy; and yet,  without any 
apparent UV, optical, or infrared 
counterpart to confirm such expectations.
Adding to the mystery, this central concentration and indeed the entire chain complex 
 has very little kinematic structure, $<$200\,km\,s$^{-1}$, making it difficult to identify any cloud rotation
 to definitively identify a discrete structure (i.e., a galaxy). The central mass concentration, requires some amount of dark matter should it be a stable structure.

The \HI\ complex, which appears to be interconnected by a tenuous `chain' of \HI, is within or at the edge of a large galaxy group, 
whose velocity dispersion is $228.6,\mathrm{km}\,\mathrm{s}^{-1}$ and with a corresponding dynamical mass of $10^{13.5}$\,$M_\odot$.
It is located some $\sim$400 kpc in projection from the group center, a gas-poor compact group of early-type galaxies. Indeed, there does not seem to be any \HI\ emission in the vicinity of the cloud complex, a relatively gas-poor region except for the complex reported here.   
Its `chain' morphology   
suggests an origin and evolution that is connected to an interaction within the galaxy group, in which case we estimate an upper limit to its age of $700\,\mathrm{Myr}$. Comparing the dark cloud chain to dark \HI\ clouds in the literature, it may be a larger version of H\,1225+01, but further observations are required, notably at higher spectral resolutions, to decode the nature of this enigmatic \HI\ gas cloud complex. 

The fact that inteferometric MeerKAT was able to detect and resolve (unlike HIPASS) this extraordinary gas complex, underscores the power of MeerKAT to survey large regions, and the importance of the SKA paradigm to raise \HI\ astronomy to the next level.

\vspace{20pt}

%

\section{Acknowledgements}
We thank Steve Schneider for a thorough review of this paper, with a number of helpful suggestions that have been implemented. We thank Thijs van der Hulst, Danielle Lucero and John Hibbard for enlightening discussions.
MC is a recipient of an Australian Research Council Future Fellowship (project number FT170100273) funded by the Australian Government. T.H.J. acknowledge support from the National Research Foundation (South Africa).\\
The MeerKAT telescope is operated by the South African Radio Astronomy Observatory, which is a facility of the National Research Foundation, an agency of the Department of Science and Innovation.\\
Part of the data published here have been reduced using the CARACal pipeline, partially supported by ERC Starting grant number 679629 “FORNAX”, MAECI Grant Number ZA18GR02, DST-NRF Grant Number 113121 as part of the ISARP Joint Research Scheme, and BMBF project 05A17PC2 for D-MeerKAT. Information about CARACal can be obtained online under the URL:: \url{https://caracal.readthedocs.io}.\\
This publication makes use of data products from the Wide-field Infrared Survey Explorer, which is a joint project of the University of California, Los Angeles, and the Jet Propulsion Laboratory/California Institute of Technology, funded by the National Aeronautics and Space Administration.\\
GAMA is a joint European-Australasian project based around a spectroscopic campaign using the Anglo- Australian Telescope. The GAMA input catalog is based on data taken from the Sloan Digital Sky Survey and the UKIRT Infrared Deep Sky Survey. Complementary imaging of the GAMA regions is being obtained by a number of independent survey programs including GALEX MIS, VST KIDS, VISTA VIKING, WISE, Herschel- ATLAS, GMRT and ASKAP providing UV to radio coverage. GAMA is funded by the STFC (UK), the ARC (Australia), the AAO, and the participating institutions. The GAMA website is http://www.gama- survey.org/. Based on observations made with ESO Telescopes at the La Silla Paranal Observatory under programme ID 177.A-3016.

\software{
CARACal \citep{Jozsa2020},
equolver \citep{jozsa_equolver},
SoFiA \citep{serra_sofia:_2015},
Stimela \citep{makhathini2018},
AOFlagger \citep{offringa_aoflagger:_2010},
CUBICAL \citep{kenyon_cubical_2018},
WSClean \citep{offringa_wsclean:_2014}, 
iDaVIE \citep{Jar21}
}

%

\newpage
\bibliographystyle{aasjournal}
\bibliography{darkchain}

%

\appendix
%
\section{MeerHOGS survey layout and observational parameters}\label{appendix_layout}

\begin{figure}
\begin{center}
\hspace*{-20pt}
\includegraphics[width=0.55\textwidth]{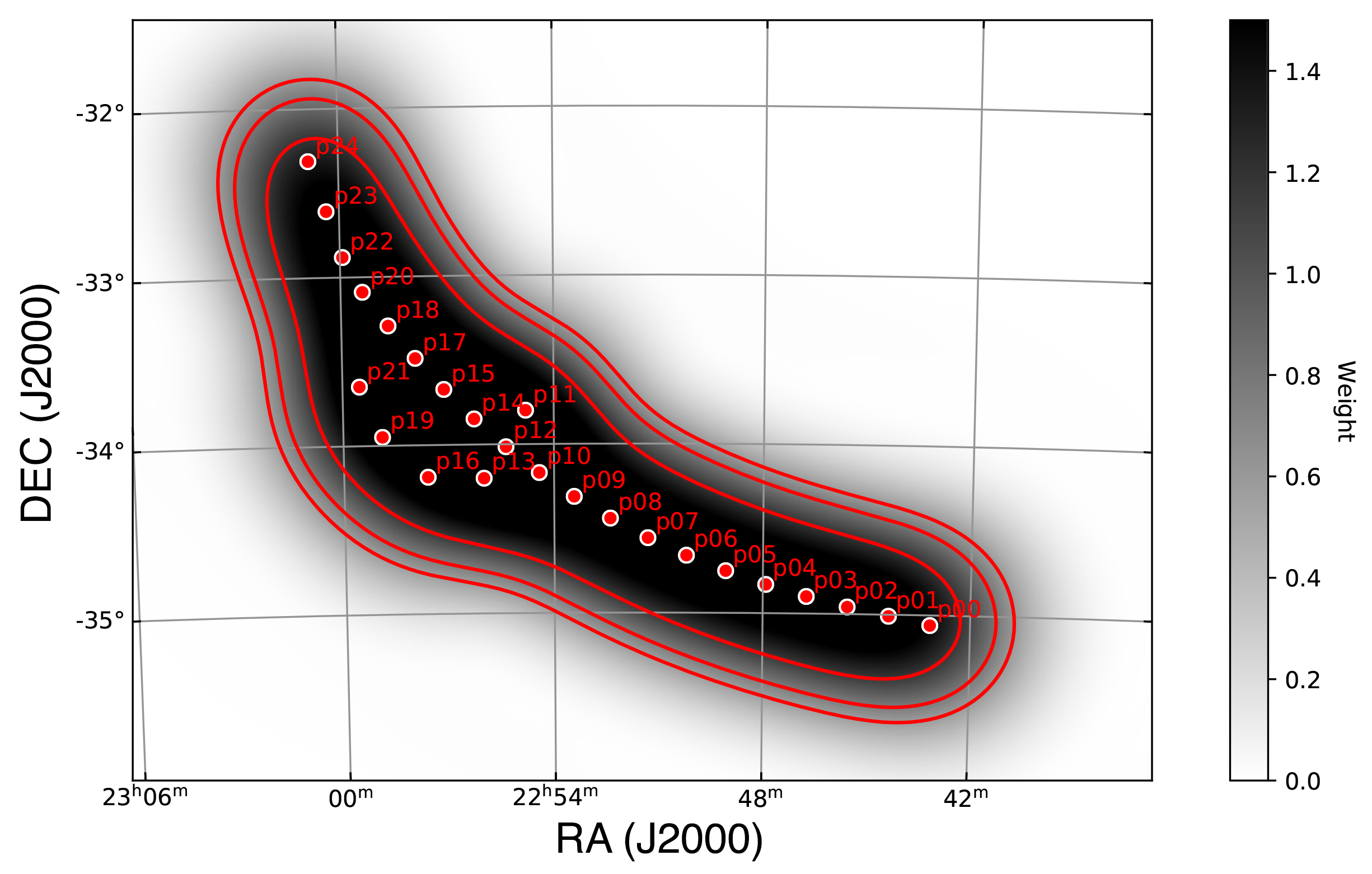}
\end{center}
\vspace{-15pt}
 \caption{Observation footprint of the MeerHOGS field.  The red points are the MeerKAT beam centers, with a total of 25 pointings.  The combination of the ($\sim$Gaussian) beams creates the footprint, where the greyscale shows the accumulated coverage (weight), with the highest coverage ($\sim$2 hours) in the central region, and the equivalent of one hour toward the edge (2nd contour).   
}
 \label{fig:mosaic_layout}
\end{figure}

%

\subsection{MeerHOGS}
Different to traditional blind surveys, MeerHOGS pointings are informed by the distribution of galaxies in a particular redshift range, thus optimised to detect galaxies residing within large-scale environments (see Cluver et al. in prep. for further details). The current study is based on a pilot survey, the effective area is closer to $\sim$7\,deg$^2$, achieving
the target HI sensitivity of 0.5\,mJy over 209\,kHz.
A study of the 1.4\,Ghz continuum sources from MeerHOGS data was carried out by Yao et al. (submitted).

Illustrated in Fig.~\ref{fig:mosaic_layout}, 
we positioned 20 pointings along the
filament, and a further 5 pointings in the area of the dense galaxy group. The total integration time for each pointing
was $\sim$30\,min. Due to the overlap between the pointings, this corresponds to a co-added equivalent integration time of $\sim$2\,hours at the centre of the target structure and an equivalent integration time of approximately an hour along the filament (second contour in Fig.~\ref{fig:mosaic_layout}).

%

\subsection{MeerKAT Observations}
The MeerHOGS observations took place in four epochs in the second half of May 2019, with a duration of 4 -- 4.5~h per observing run.  The data were observed using the MeerKAT SCARAB Correlator in 4K mode.  The total bandwidth was $\sim$800\,MHz, with 4096 channels of width 209\,kHz, corresponding to 45.4\,km\,s$^{-1}$ at z\,$\sim$\,0.02962.
These observations were part of the MeerKAT Early Science program, which used a simpler 4K correlator. In the future, the full 32K channel mode will be available to significantly improve the velocity resolution.

In each observing run, each pointing was visited with an on-source integration time of 15\,min. Between the target observations the gain calibrator J2302-3718 was visited for 2 min and the bandpass-- and flux calibrator PKS\,1934-63 was observed three times for 10 min in each observing epoch. The total observing time was 16.5\,h, with 13.2\,h being spent on-source to create the mosaic in Fig.~\ref{fig:mosaic_layout}.

The achieved angular resolution (HPBW) in the continuum image after mosaicing is 
 $13.\!\!\arcsec5 \,\times\, 13.\!\!\arcsec5$, which is uniform throughout the MeerHOGs region, while the angular resolution of the line data cube is $34.\!\!\arcsec4 \,\times\, 34.\!\!\arcsec4$; 
further observation details given in Table~\ref{tab:obs_info},
and detailed \HI\ data reductions are below.

%

\subsection{Data reduction: CARACal}
Only part of the data were transferred from the SARAO archive for the \HI\ data reductions, using the parallel-hand correlation products only. We used the frequency range from 1319.8 to 1517.1\,MHz to both avoid RFI-dominated frequencies as well as to keep the data volume manageable. The primary data reduction was conducted using the platform-independent {\sc CARACal} pipeline\footnote{\url{https://caracal.readthedocs.io}} pipeline \citep{Jozsa2020}. The data reduction took place on workstations typically using 36 computing cores and 300\,GB of memory. {\sc CARACal} makes use of the {\sc Stimela} {\sc Python} framework, which combines a large variety of available data reduction software through containerisation and a common {\sc Python} interface \citep{makhathini2018}. Unless otherwise stated, all steps described in the following were made from within the pipeline and hence the interface. We performed a standard interferometric data reduction including a self-calibration and subsequent mosaicing.

After downloading the data, a data set containing only the calibrator observations (PKS\,1934-63 and J2302-3718) was constructed. For the resulting data set the first dump after each setup change (``quack-flagging'') and autocorrelations were flagged. The data were flagged for shadowing, then for RFI using {\sc AOFlagger} \citep{offringa_aoflagger:_2010}. For the cross-calibration baselines above 150~m only were used, to perform a delay-, bandpass-, and gain calibration using the primary calibrator (PKS\,1934-63). The gains were then transferred to the secondary calibrator (J2302-3718), to apply a combined gain- and flux calibration, flag the data based on the solutions, then apply a combined gain- and flux calibration again. The resulting calibration tables were transferred to the target observations. Following that, the target data were flagged again for setup changes, autocorrelations, shadowing, and RFI (using {\sc AOFlagger}). 

For the continuum imaging with {\sc WSClean} \citep{offringa_wsclean:_2014} the target data were then binned by a factor of 16 to a channel resolution of 3.3\,MHz and potential \HI\ emission from the Milky Way (around 1420.4~MHz) was flagged in this data set. A Gaussian taper of 7~arcsec was applied to the data while using a Robust weighting of -1. A region of $3\,\times\,3\deg$ was imaged four times per pointing. For each imaging step, a CLEAN mask was generated using the SoFiA \citep{serra_sofia:_2015} source-finding software (using a threshold of $4,4,3.5,3.5\,\sigma_\mathrm{rms}$, where $\sigma_\mathrm{rms}$ is the locally measured rms noise), to CLEAN the data with a CLEAN cutoff-parameter of 0.2. The resulting CLEAN models were used for self-calibration with {\sc CUBICAL} \citep{kenyon_cubical_2018}, correcting the phases only of the visibilities after the first two imaging iterations, and both the phase and the gains after the third imaging iteration. Using the {\sc equolver}\footnote{\url{https://pypi.org/project/equolver/}} package (at the time of the data reduction not yet implemented in {\sc CARACal}) the resulting continuum images were convolved to a common resolution of $13.\!\!\arcsec5\,\times\,13.\!\!\arcsec5$, where $13.\!\!\arcsec5$ was the maximum $HPBW$ in any direction. The resulting images had an rms noise of $\sigma_\mathrm{rms}\,=\,18\pm 2\,\mathrm{\mu Jy}$. The images were finally mosaiced to the final continuum image, weighted by the (quadrature of) the primary beam, for which we used the prescription by \citet{mauch_128_ghz_2020}.

%

\subsection{\textrm{H}\textsl{\textsc{i}}\ Data Cubes}

Prior to line imaging, the gains obtained from the continuum self-calibration were applied to the original data with higher frequency resolution and the continuum models used in the self-calibration step were subtracted from the visibilities. For computational reasons, the data were then split into chunks of 20~MHz with an overlap of 5~MHz. For each chunk, the velocity reference frames were shifted from the (original) topocentric to heliocentric and an additional continuum subtraction was applied by fitting a 3rd order polynomial to the visibilities. The data were then imaged two times using a Robust weighting of 0 and a Gaussian taper of 10\arcsec, to create data cubes with spatial dimensions of $2\,\times\,2\deg$. 

The data were CLEANED in a first iteration using {\sc WSClean} automasking (with a CLEAN threshold of $10\,\sigma_\mathrm{rms}$), to use again a 
CLEAN mask as calculated using the SoFiA source-finding software with a clip level of $4\,\sigma_\mathrm{rms}$ and a CLEAN cutoff level of $0.5\,\sigma_\mathrm{rms}$. For each frequency range, the resulting cubes were convolved to the same resolution using {\sc equolver} and then mosaiced, again weighted using the (quadrature of) the primary beam. To achieve a coherent data cube spanning the whole frequency range the data cubes were then combined again, using an interpolation in the overlapping frequency/velocity ranges.  A final cube was thus created, dimensions 5.7\degr$\,\times$\,4.9\degr, and 479 velocity channels.

%

\subsection{Source finding and image-domain continuum subtraction}\label{sect_source_finding}
To identify \HI\ sources and to improve the continuum subtraction, we once again employed SoFiA. For the source detection, the images were smoothed only in the spatial domain, and not the frequency domain, using 4 Gaussian filters with sizes 0, 3, 6, and 9 voxels. A mask was created by clipping the emission in each resulting convolved cube at a $4-\sigma_\mathrm{rms}$-level and combining the resulting mask. Subsequently, sources were identified as voxel-complexes (or ``islands") in the mask and a reliability calculation was applied to reject islands/sources whose statistical properties indicate a spurious detection (for details see \citealt{serra_sofia:_2015}). Using our initial continuum subtraction in the visibility-domain, we did not exclude channels with emission, as with the large field-of-view of MeerKAT the number of channels removed from the calculation of the continuum data cube can become very large. However, this leads to a slight over-subtraction in channels around bright \HI\ sources.

To mitigate any over-subtractions, we applied an additional continuum modeling to the data cube. A second-order Savitzky-Golay \citep{savitzky_1964} filter of the length of 51 channels was applied to the data cube, to create a residual continuum cube. Doing so, we interpolated (the continuum-) emission along the frequency axis at the positions of detected sources. The resulting continuum cube was subtracted from the original data cube, resulting in our final data cube, on which we ran SoFiA one final time. In the final run of SoFiA a source list was generated. The masks were inspected and those occasions identified where potential residual flux from strong continuum sources could mimic an \HI\ source. Those were flagged and removed from the subsequent analysis. In total, 196 sources were detected. No strong continuum source was found in the vicinity of the constellation described in this paper and we did not notice any strong or even moderate signatures of RFI in the channel ranges discussed in this paper. We hence exclude the possibility that our observations reflect any type of measurement error.

%

\section{VR Visualisation of the dark cloud chain}\label{appendix_b}

We deployed a new cutting edge visualisation tool to investigate the dark cloud chain.  Designed specifically to work with spectral-imaging cubes, \IDAVIE\ (Immersive Data Visualization Interactive Explorer\footnote{https://idavie.readthedocs.io/en/latest/ and Jarrett et al. 2020}) renders cubes in 
a room-scale 3D Virtual Reality (VR) space, 
where the user can
intuitively view and uniquely interact with their data cube \citep{Jar21}.  The software suite is designed to work with FITS cubes and products from the SoFiA\footnote{https://github.com/SoFiA-Admin/SoFiA} \citep{serra_sofia:_2015}
source finding package, including the mask cubes and extracted source lists. \IDAVIE\ also has the capability of uploading redshift catalogues, which we utilized to confirm cross-matches between the \HI\ and optical detections in the MeerHOGS region.

\begin{figure*}[ht!]
\includegraphics[width=1\textwidth]{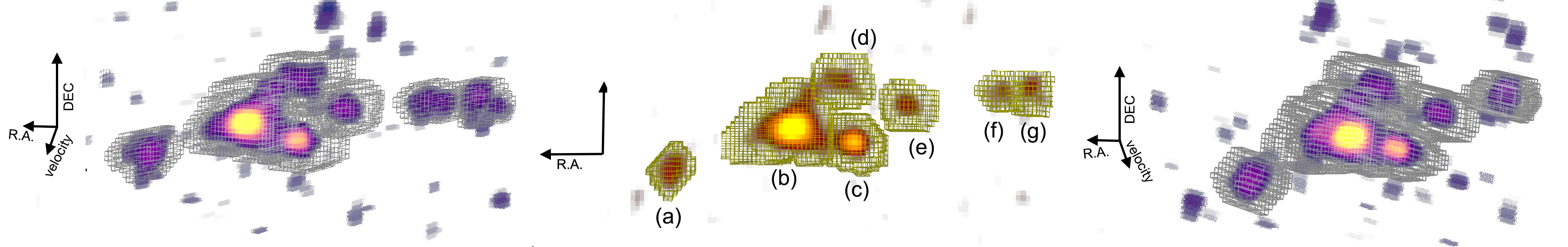}
\caption{Three orientation views of the dark cloud chain using Virtual Reality and the \IDAVIE\ application.  The 3D rendering is seen from the lens of the VR user.
The \HI\ gas distribution is shown with a color transform (``plasma", "gist-heat", ``plasma" for the three views, respectively). The grid mesh demarcates voxels that are identified as having emission associated with the seven sources of the dark chain. The diagram shows that the chain is stretched along the E-W spatial plane, while relatively flat (only 2 to 3 channels) along the radial/velocity axis.
}
\label{fig:idavie-horiz}
\vspace{-10pt}
\end{figure*}

The dark cloud chain of \HI\ emission is composed of several close peak concentrations, essentially blended into one object by the initial SoFiA extraction.  We used \IDAVIE\ to identify and separate the individual concentrations through a voxel editing process in which the mask was user-edited in 3D, based on the spatial and kinematic distribution of the \HI\ gas.  The result of this human-pattern-recognition source deblending is shown in Fig.~\ref{fig:idavie-horiz}, which shows a 3D rendering of the dark cloud chain as seen through the lens of the VR user.  The \HI\ emission is transformed to a color scale (here, we are showing two different examples).  Although limited by the 2D flat page, these three views demonstrate the 3D structure of the dark cloud chain, characterized by dense gas concentrations stretched into a filamentary cloud complex, and with very little radial-velocity kinematics (2 to 3 channels) in the orthogonal direction.  Finally, the edited mask is depicted with a transparent mesh grid, overlaid on the \HI\ emission.  Visually discerning which voxels need editing, the mask value is tagged to the identified source, which may then be used to extract the source and characterize it.

As first presented in Fig.~\ref{fig:4plot} and Table 2, the sources are labeled (a) to (g).   The mask determines which voxels are used to compute the source characteristics (e.g., systemic velocity, spatial coordinates, integrated flux, etc).  This method of dividing the chain into pieces based on the peak emission and extent is only an approximate deblending.  It is inevitable that some emission will bleed over into neighboring sources; moreover, 
it is not clear if all the sources (or some combination thereof) are discrete or part of a larger diffuse and filamentary cloud.  Nevertheless, the most massive of the group of seven, labeled (b), is both central and appears to be a discrete galaxy-sized object.  The \IDAVIE\ extraction of this central source is about as good as it can be done given the angular resolution (34\arcsec HPBW) and kinematic resolution (45 km\,s$^{-1}$) of the \HI\ cube mosaic.

%

\vspace{+10pt}
\section{HIPASS Spectrum of the Dark Cloud Chain}\label{appendix_c}

HIPASS carried out a systematic \HI\ survey of the
southern hemisphere using the Parkes 64-m dish and multi-beam spectrometer \citep{barnes_hi_2001}.  The velocity range of the survey extends to $12700\,\mathrm{km}\,\mathrm{s}^{-1}$, but rapidly decreases in completeness beyond $\sim$4800\,km\,s$^{-1}$.  Nevertheless, massive \HI\ sources may be detected at more distant velocities; since the dark cloud is at v$_{opt}$\,$\sim 8900\,\mathrm{km}\,\mathrm{s}^{-1}$, we can at best expect a weak detection (the source does not exist in any catalogued HIPASS or HICAT listings).

We do, however, locate the source in region H187 of HIPASS. Using an improved data cube created for the HICAT (catalogue) processing \citep{Meyer04, Zwaan04},  
we extracted a spectrum centered on Source (b) (see Table~\ref{tab:HIprop}), both its spatial and velocity intensity peaks and detected a source at its location and velocity. 
To improve the signal-to-noise, we averaged the \HI\ 
surface brightness 
across $\pm$3 pixels in the spatial axis, or a region 12\arcmin\ in width (the pixel scale of the cube is 4\arcmin, while the spatial resolution or beam is about 15.5\arcmin, and hence native pixels are highly correlated).  

\begin{figure*}[b!]
\includegraphics[width=0.5\textwidth]{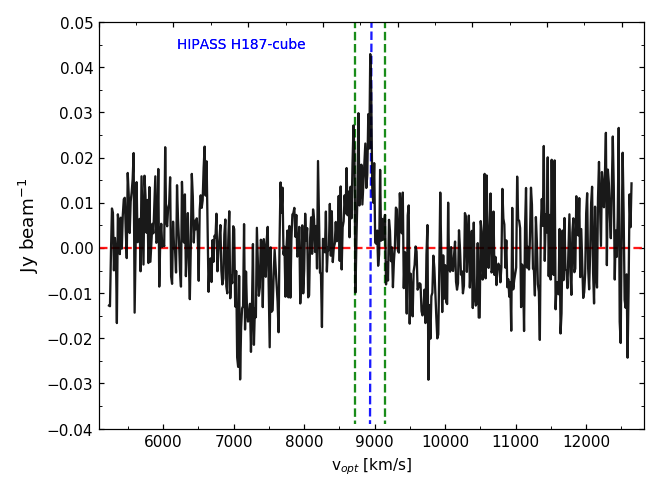}
\includegraphics[width=0.5\textwidth]{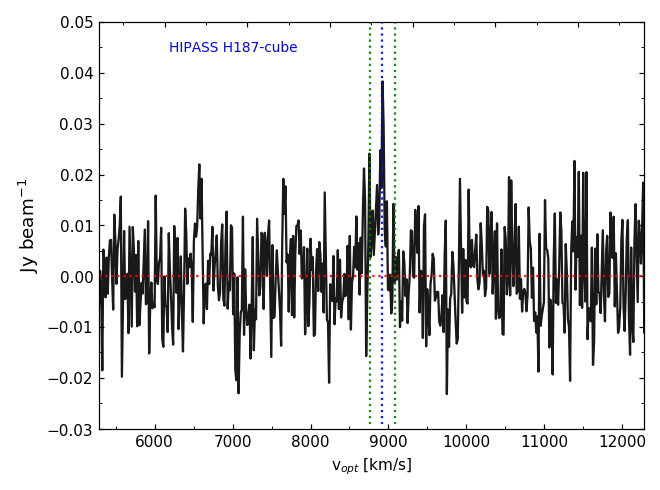}
\caption{\HI\ spectrum of the dark cloud chain as measured using the HIPASS cube H187 (wide-band).  The variation in the underlying continuum (left) has been fit and removed (right panel).  The source is clearly detected at the spatial --- RA,DEC[deg J2000]\,=\,344.080, -33.958 --- and velocity coordinates of the massive \HI\ Source (b) (see blue vertical line at 8886\,km\,s$^{-1}$).  The green lines demark the velocity limits for the spectrum integral.
}
\label{fig:hipass}
\vspace{-10pt}
\end{figure*}

The source is clearly detected in the spectrum.  There is, however, substantial variation in the background continuum levels due to the well-known `standing wave' ripple phenomenon inherent to single dish radio telescopes \citep{briggs_driftscan_1997}.  Although there may be more optimal methods for removing said artifacts from the data (e.g., in the uv-domain; see \citet{pop08}), we have simply used a cubic-spline fit to the spectrum to remove the first-order variations.  We note that our goal here is to extract a coarse integrated flux of the source to compare with our MeerKAT measurements.

We experimented with fitting high-order polynomials and cubic splines with similar results.  Fitting a spline3 to the underlying continuum (with the source itself masked from this fitting procedure to avoid over-fitting and removing the source itself), the resulting spectrum is shown in Fig.~\ref{fig:hipass}, before and after ripple correction; the channel width is 13.2\,km\,s$^{-1}$. A zoomed velocity range is shown the main text, centered on the cloud velocity, 
Fig.~\ref{fig:hipass2}. 
 It is evident the large variations are removed for the most part, leaving a faint source detection at center. The channel-to-channel scatter in the background continuum (excluding the source) is 8.95 mJy, which we use to estimate the measurement signal-to-noise.

The peak emission is at a velocity of $\sim 8886 \mathrm{km}\,\mathrm{s}^{-1}$
with a flux density of 
0.038\,Jy/beam at the S/N=4.2 level. 
This peak corresponds to the peak velocity of Source (b) (see blue line in Fig.~\ref{fig:hipass}), which lends confidence to a real detection in the HIPASS cubes.  
Integrating the spectrum channels across $\pm$150\,km\,s$^{-1}$ (a wide enough range to capture all of the gas in system; see green lines in Figure),
centered on the cloud systemic velocity (blue line), the integrated flux is 2.9$\pm$0.6\,Jy\,km\,s$^{-1}$.  

For the MeerKAT observations, we measure the integrated flux of Source (b) to be 1.4\,Jy\,km\,s$^{-1}$, and of the entire complex to be $\sim$2.8\,Jy\,km\,s$^{-1}$.  Hence, the single-dish HIPASS results are 
are consistent with the total flux of the dark cloud chain, given the large Parkes Telescope beam ($\sim$15.5\arcmin) and excellent spatial coverage.  These results lend confidence to a detection in the HIPASS survey, and conversely, to validation of the MeerKAT measurements.
\end{document}